\documentclass[11pt]{article}
\usepackage{makeidx}
\usepackage{multirow}
\usepackage{multicol}
\usepackage[dvipsnames,svgnames,table]{xcolor}
\usepackage{graphicx}

\usepackage{ulem}
\usepackage{hyperref}
\usepackage{amsmath}
\usepackage{amssymb}
\author{Gianni}
\title{}
\usepackage[paperwidth=595pt,paperheight=841pt,top=70pt,right=70pt,bottom=70pt,left=70pt]{geometry}

\makeatletter
	{\par\setlength{\parindent}{#3}
	\setlength{\leftmargin}{#1}       \setlength{\rightmargin}{#2}%
	\advance\linewidth -\leftmargin       \advance\linewidth -\rightmargin%
	\advance\@totalleftmargin\leftmargin  \@setpar{{\@@par}}%
	\parshape 1\@totalleftmargin \linewidth\ignorespaces}{\par}%
\makeatother

\begin{document}

\begin{center}
\textbf{{\Large    Title   The $\kappa$-model, a minimal model alternative to dark
matter. Application to the galactic rotation problem}}
\end{center}

\begin{center}
{\Large G. Pascoli }
\end{center}

\begin{center}
{\Large emails : pascoli@u-picardie.fr; sciences.univers@gmail.com}
\end{center}

\begin{center}
{\Large Facult\'{e} de sciences, D\'{e}partement de physique}
\end{center}

\begin{center}
{\Large Universit\'{e} de Picardie Jules Verne (UPJV)}
\end{center}

\begin{center}
{\Large 33 rue saint Leu,   Amiens, France}
\end{center}

\textbf{{\Large Abstract}} The determination of
the velocities, accelerations and the gravitational field intensity at a given
location in a galaxy could potentially be achieved in an unexpected manner with
the environment of the observer, for instance, the local mean mass density in the
galaxy. This idea, mathematically supported by the asymmetric distance
concept, is illustrated here by a study regarding the rotation of spiral
galaxies. This suggestion is new in the astrophysics field (in the
following, it is called the $\kappa$-model) and could help to mimic the main effects seen in modified Newtonian dynamics (MOND) theory, modified gravity (MOG) models, or other related
models built with the aim of eliminating dark matter that are already
well-established theories. Thus, starting from  selected examples of galaxies, in sections 5, 6 and 7 we show that there is an equivalence between MOND and the $\kappa$-model. In particular, on the opposite side, we have the speculative nature of the dominant paradigm, the elusive dark matter, a
matter whose properties always remain undefined despite intense
theoretical, experimental and observational efforts for over 50 years.

\vspace{10pt}


\section{Introduction}

Alternative theories to dark matter attribute the flatness of the
galactic rotation curve either to a departure from the law of inertia (modified Newtonian dynamics (MOND))
in the very low acceleration regime (Milgrom, 1983, 2009) or to a failure
of Newtonian gravity at a large scale, emulated by a variable
gravitational constant $G$ (Moffat, 2006, 2008), and neither invoke an
additional very large quantity of dark matter. Many other models that do not involve dark matter have also been proposed (Capozziello and De Laurentis M., 2012; O'Brien and Moss, 2015; Ludwig, 2021). One of them, conformal gravity (CG), has generated extensive interest since it claims to explain the flat rotation curve of galaxies without dark matter. The majority of the CG is based on the Mannheim-Kazanas vacuum solution of the CG field equations for a static, spherically symmetric spacetime (Mannheim and Kazanas, 1989). Nevertheless, the conformal equivalence of the Mannheim-Kazanas and Schwarzschild–de Sitter metrics strongly suggests that the prediction of a flat rotation curve for the galaxies by this type of model may be a
gauge artifact since performing a similar analysis in the
Schwarzschild–de Sitter metric yields rotation curves without any flat region (Hobson and Lasenby, 2021).

Negative masses have also been invoked (Farnes, 2018; Benoit-Lévy and Chardin, 2012; Manfredi, Rouet, Miller and Chardin, 2018). However, a careful analysis of the negative mass concept identifies a number of incompatibilities with existing observations (Socas-Navarro, 2019). Regardless, we can ask what is the benefit of replacing an unobserved exotic entity, i.e., dark matter, with another unobserved, even more
exotic entity, for instance, a fluid composed of negative mass. On the other hand, a correlation between the radial acceleration traced by rotation curves and that predicted by the observed distribution of baryons has been reported (McGaugh, Lelli and Schombert, 2016). This radial acceleration relation seems tantamount to a natural law for rotating galaxies and is not compatible with the hidden presence of a large amount of exotic mass.

There are other channels that have yet to be explored. The assumption developed in the present paper, namely, that both gravitational field intensity and the
measurement of the velocities and accelerations of a particle could depend on
the mean mass density estimated in a given location, has not been considered thus far. However,
this idea, even though very simple, offers new perspectives because of its originality and its various consequences (Pascoli, and Pernas, 2020). The key strengths of this
concept are as follows:

\begin{enumerate}

	\item The basic elements of this assumption are very natural and reside within the framework of the
Newtonian mechanics. Let us recall that Newtonian mechanics is relevant only when the velocities are low and the
gravitational field is weak, but this is the case here, even though the model
can still be derived from a relativistic context\footnote{See(Pascoli, and Pernas, 2020)
for an attempt to develop the $\kappa$-model in a relativistic context.}. The
concept, hereinafter named the $\kappa$-model, postulates an enhancement, resp. a
diminution, of the self-gravity in the regions where the
mean mass density is weak, resp. high, in a galaxy. An associated effect is
that the measured distances become apparent and asymmetric. This statement
is not explicit in the context of either the MOND paradigm (Milgrom, 1983, 2019)
or modified gravity (MOG) theories (Moffat, 2006, 2008). In addition, the $\kappa$-model remains in the strict framework of a preserved Newtonian law of gravity,
at least from a formal point of view. More generally, all the physical laws,
locally expressed, remain unmodified in the $\kappa$-model.

	\item The persistence of a spiral substructure in "grand design" galaxies and the
flatness of the rotation curves are shown to be interrelated in an unexpected
manner.

	\item Some problems such as the bullet cluster that are seemingly difficult to solve in the
framework of the various alternatives to the dark matter paradigm (for
instance, MOND) could perhaps be naturally explained under this framework. This is because
gravity, where it acts, is now modulated by the local mean density.
Another issue is related to the accelerating expansion of the Universe
(Frieman, Turner and Dragan, 2008; Weinberg, 2008). However, this problem could also be solved in the
framework of the $\kappa$-model, eventually leading from two distinct paradigms
(dark matter and dark energy) to a single one. All these important issues,
where a possible density-dependent aspect of the gravitational force is put in
evidence, are considered later.

\end{enumerate}

\section{The \texorpdfstring{$\kappa$-model}{Lg}}

The equations of this framework are
the usual Newtonian (nonrelativistic) equations, except they are weighted by a coefficient $\kappa$, as explained immediately below \footnote{The relativistic
expression is not useful here, even though the relativistic transcript would be
easy to form in the kappa model framework.}. For a system composed of
$N$ identical particles of mass $m$ (index $i=1..., N)$, we write

\begin{equation}
\frac{d}{dt}\left({\kappa{}}_i[\bar{\rho{}}]\frac{d{\boldsymbol{\sigma{}}}_i}{dt}\right)=-Gm\sum_{j=1,\
\
j\not=i}^N\frac{{\kappa{}}_i[\bar{\rho{}}]({\boldsymbol{\sigma{}}}_i-{\boldsymbol{\sigma{}}}_j)}{{[{\kappa{}}_i[\bar{\rho{}}]\
\left\Vert{}{\boldsymbol{\sigma{}}}_i-\boldsymbol{\sigma{}}_j\right\Vert{}]}^3}
\end{equation}

\subsection{A formal deduction of equation (1) from the gravitational Newtonian law}

  Here, we start with the Newton gravitational law, as it is confirmed to be accurate at the local level (the solar system). Let us define two very distant particles $M$ and ${M}'$ with respective masses $m$ and ${m}'$, denoted by the vectors $\bf{R}$ and ${\bf{R}}'$ (arbitrary origin). We write formally for the particle $M$, assuming that the gravitational constant $G$ is universal

\[
\frac{d\bf{P}}{dt}=-Gmm'\frac{(\bf{R}-\bf{R}')}{{\ \left\Vert{}\bf{R}-\bf{R}'\right\Vert{}}^3}
\]

{\raggedright where ${\bf{P}}$ is the momentum of particle $M$, and we can write a similar equation for ${M}'$. At this stage, there is no observer (no reference frame) to measure the
kinematic quantities. As is well known in mechanics, without a
reference frame, both the velocity and acceleration are undetermined. Likewise,
$\frac{d\bf{P}}{dt}$ cannot be added to $\frac{d\bf{P}'}{dt}$ apart from in a purely formal manner. In addition, in the $\kappa$-model, the norms of $\bf{R}$ and $\bf{R}'$ are now themselves left undetermined. Thus, in
To solve this pair of equations, we must choose a representation: a
fictitious inertial observer $A$ is assumed to be located near $M$ (resp.
$B$ near ${M}'$), each of these observers is equipped with a scale factor
$\kappa{}$ (resp. $\kappa{}'$). Then, $A$ locally expresses ${\bf{R}},{\bf{R}'}$ and ${\bf{P}}$:
${\bf{R}}\longrightarrow{} \kappa{}{\boldsymbol{\sigma{}}},\
{\bf{R}'}\longrightarrow{}\kappa{}{\boldsymbol{\sigma{}}}'$ and ${\bf{P}} \longrightarrow{}   \kappa{}(m\frac{d{\boldsymbol{\sigma{}}}}{dt})$, and $B$ does the same thing
but with a scale factor $\kappa{}'$. We can note that there exists some -- even though obviously  very
remote -- mathematical analogy with quantum mechanics when we pass from the operational
notation to the Schr\"{o}dinger representation for the position $\bf{R}$ and
momentum $\bf{P}$ of a point particle ($\hat{\bf{R}}\longrightarrow{}\bf{R},\
\hat{\bf{P}}\longrightarrow{}-i\hslash{}\boldsymbol{\nabla{}}$).}

 The scale factor (the mathematics) is then linked to the mean density by equation (2) introduced below (the physics). Eventually, we assume that the fictitious (inertial)
observers $A$ and $B$ are motionless with respect to each other, an assumptions that is easily verified by
spectroscopic measurements. Another deduction of equation (1) issued from the variational principle is given in appendix A.

\subsection{\texorpdfstring{coefficient $\kappa$}{Lg}}

In equation (1), the coefficient $\kappa{}$ is no longer a constant (equal to $\kappa_E$) as in the basic Newtonian equations. In contrast, $\kappa{}$ is now defined as a functional of the mean local density
$\bar{\rho{}}$\footnote{Note that the densities in equation (2) are the
densities estimated by the same observer or, more precisely, that $\rho_0$ and $\bar{\rho}$ are simultaneously measured by this observer.}. Obviously, at the local scale from our perspective,
i.e., at the scale of the Earth or the solar system, we can take ${\kappa{}}_i=\kappa_E=Const,\ \forall{}i$ and analyze a group of particles
contained in the solar system (planets, asteroids, etc.); these equations exactly express the
usual Newtonian law.

By considering the fact
that $\kappa{}$ is assumed to be a smooth linear function and that $\bar{\rho{}}$ is an exponential function in a typical galaxy (Binney and Merrifield, 1998), it appears intuitive to impose a natural and simplest form, i.e., a logarithmic relationship between $\kappa{}$ and $\bar{\rho{}}$. Let

\begin{equation}
\frac{\kappa{}[{\rho{}}_0]}{\kappa{}\left[\bar{\rho{}}\right]}=1+Ln(\frac{{\rho{}}_0}{\bar{\rho{}}})
\end{equation}

{\raggedright where the index $0$ labels the maximum value of the density distribution
($Ln$ is the symbol for the Napierian logarithm). This law is assumed to be universal
and available for any galaxy.}

At this level, the system of equations (1) and (2) can be admitted from the outset as a
postulate of the model, and we could simply end here, leaving the
deduction from first principles aside. However, the reader can refer to
appendix A for a synthetic demonstration of the dynamics formula represented by equation (1).

Let us examine equation (1). The inertia term is modulated by the factor
$\kappa{}$ (as in MOND). On the other hand, from the right-hand side, we can
see that wherever the mean density is high (resp. low), the gravitational
"sensation" between two masses is low (resp. high). Two masses are more
strongly gravitationally linked with each other in the outer regions (low
density, low $\kappa{}$) of a galaxy than in the inner regions (high density,
high $\kappa{}$). This bears some resemblance to electromagnetism in a
situation where two charged particles are placed in a medium of relative
dielectric permittivity  ${\epsilon{}}_r$ different from unity
(${\epsilon{}}_r>1$). The electric field between these charges is lower than that of vacuum. However, this analogy must not be taken too far. There exists an
important difference: no background medium exists in the $\kappa$-model, and the coefficient $\kappa{}$ induces no refractive
effect. The light still propagates in a straight line with the same speed
$c$ for all frequencies, and $\kappa{}$ is frequency-independent, contrary to
$\ {\epsilon{}}_r$, which depends on the frequency. Likewise, the trajectory of
any free particle is rectilinear in the $\kappa$ model (when the gravitational
force is eliminated).

\section{Computational details}

Equation (1) seems to be "easily" treated by
any standard and well-known method of smoothed particle hydrodynamics (SPH), for instance, by
using a code available online. Unfortunately, these codes are most often based on the default 4th order Runge-Kutta algorithm. In the present situation, the process of solving is more complex given that the factor $\kappa{}$ is now a functional of the mean density. Then, we are facing a self-consistent problem, and a more stable numerical algorithm than the 4th order Runge-Kutta order method must be used, even though using a higher order Runge-Kutta algorithm considerably increases the CPU runtime.

Thus, to solve the system of equations (1) and (2), an SPH code based on a 6th order Runge-Kutta ordinary differential equation (ODE) algorithm in MATLAB was used throughout. This package is currently
implemented on an SGI Altix UV 100 (MatriCS Platform) at UPJV.

A damping term is added to each equation of the system of equations (1) in
order to simulate a radial pseudoviscosity effect when the density is larger
than a fixed threshold (the inner region). This term has the simple form

\begin{equation}
-{\alpha{}}_i[{\kappa{}}_i\frac{d{{\boldsymbol{\sigma{}}} }_i}{dt}.
\frac{{{\kappa{}}_i{\boldsymbol{\sigma{}}}}_i}{\left\Vert{}{{\kappa{}}_i{\boldsymbol{\sigma{}}}}_i\right\Vert{}}]\frac{{{\kappa{}}_i{\boldsymbol{\sigma{}}}}_i}{\left\Vert{}{{\kappa{}}_i{\boldsymbol{\sigma{}}}}_i\right\Vert{}}
\end{equation}

{\raggedright where the damping coefficient
${\alpha{}}_i=\alpha{}\sqrt{\frac{4GM}{{\left\Vert{}{{\kappa{}}_i{\boldsymbol{\sigma{}}}}_i\right\Vert{}}^3}}$
 ($M$ is defined in the following, and $\alpha{}$ is a numerical coefficient,
possibly adjusted, which we have taken equal to $1$)}\footnote{From a physical
point of view, we can still imagine that when the system of particles shrinks,
the energy of the free fall is rapidly transformed in infrared radiation
which are on the spot evacuated from the system. }.

The characteristic time is taken to be equal to the free fall of an
outer particle, $\sim{}{10}^8$  years (reference unit taken for the time in
figures 1, 3, and 6). We start with an initial discoidal configuration of radius
$\sim{}10\ r_G$, where ${r}_G\sim{}10\ kpc$ (reference unit taken for the
distances in figures  1, 3, and 6), this radius is estimated by a terrestrial
observer provided with a coefficient $\kappa{}={\kappa{}}_{E}$. The total
mass of gas $M$ is set to be equal to ${10}^{44}\ g\ $(this mass is only
baryonic in the $\kappa$-model). The initial disk configuration of thickness $\sim{}0.6\ kpc,$ is assumed to be cold and homogeneous. The mean density $\bar{\rho{}}$ is automatically recalculated after 10 time steps. This means
that the density is obtained by counting the number of particles of mass $m$ in
a volume of ${(0.6\ kpc)}^3$. Eventually, an initial low shear velocity
field of the form $v_x=$  10 $\frac{{{\kappa{}}_{E}\sigma{}}_x}{10}$
${{(\kappa{}}_{E}\sigma{}}_x\ $in  $kpc$ and $v_x$  in $km/s$)
is assumed to pervade the disk. The coefficient $\kappa$    is a
constant $={\kappa{}}_{ini}$ in the initial, assumed to be homogeneous, distribution of baryonic
matter and is, for example, $\frac{{\kappa{}}_{ini}}{k_{E}}\sim{}0.15\ $, which appears to be a
reasonable value according to trial. This is for the physics content; then, the
equations are properly normalized.

The Cartesian coordinates $(x,y)$ are used. An initial uniform
distribution of $10\ 000=100\times{}100$ of identical particles of mass $m$ are selected\footnote{The study should be resumed with a larger number of
particles, but the excessive CPU time due to the self-consistent process
currently precludes this.}. To distinctly exhibit the
asymmetric substructures, the quantity reported in figures 3.a, 3.b
and 6.a, 6.b is not directly $\bar{\rho{}}\left(x,y\right)$ but rather the
difference

\begin{equation}
\delta{}\bar{\rho{}}(x,y)=\bar{\rho{}}(x,y)-\bar{\rho{}}\left(\sqrt{x^2+y^2}\right)
\end{equation}

{\raggedright where $\bar{\rho{}}\left(\sqrt{x^2+y^2}\right)$ is the density
averaged over the polar angle. The latter gives a circular distribution in the
galactic plane. Eventually, only one particle out of 10 is reported in the
figures to clearly distinguish the substructures.}

\section{Results}

When the simulation is running, the disk rapidly shrinks by
self-gravity, and a weak spiral substructure naturally quickly appears, just
after $\sim{}6\ {10}^8$ years\footnote{This characteristic time is almost twice
as long in the dark matter model, where the shrinkage is lower. The galaxies
form much more rapidly with the $\kappa$-model than with the dark matter paradigm.}. This weak substructure is shown in figure 1,
with the axisymmetric background excluded. In figure 2, the trajectories of some
individual particles are also displayed, and we can observe a rapid and
chaotic falling of these particles toward the center\footnote{The irregularities seen along on the
trajectories are due to close encounters between particles over the free
fall process. During this process, the particles that are ejected from the
galaxy are no longer taken into account.} and the
consecutive circularization of the orbits.

\begin{center}
\includegraphics[height=200pt, width=200pt]{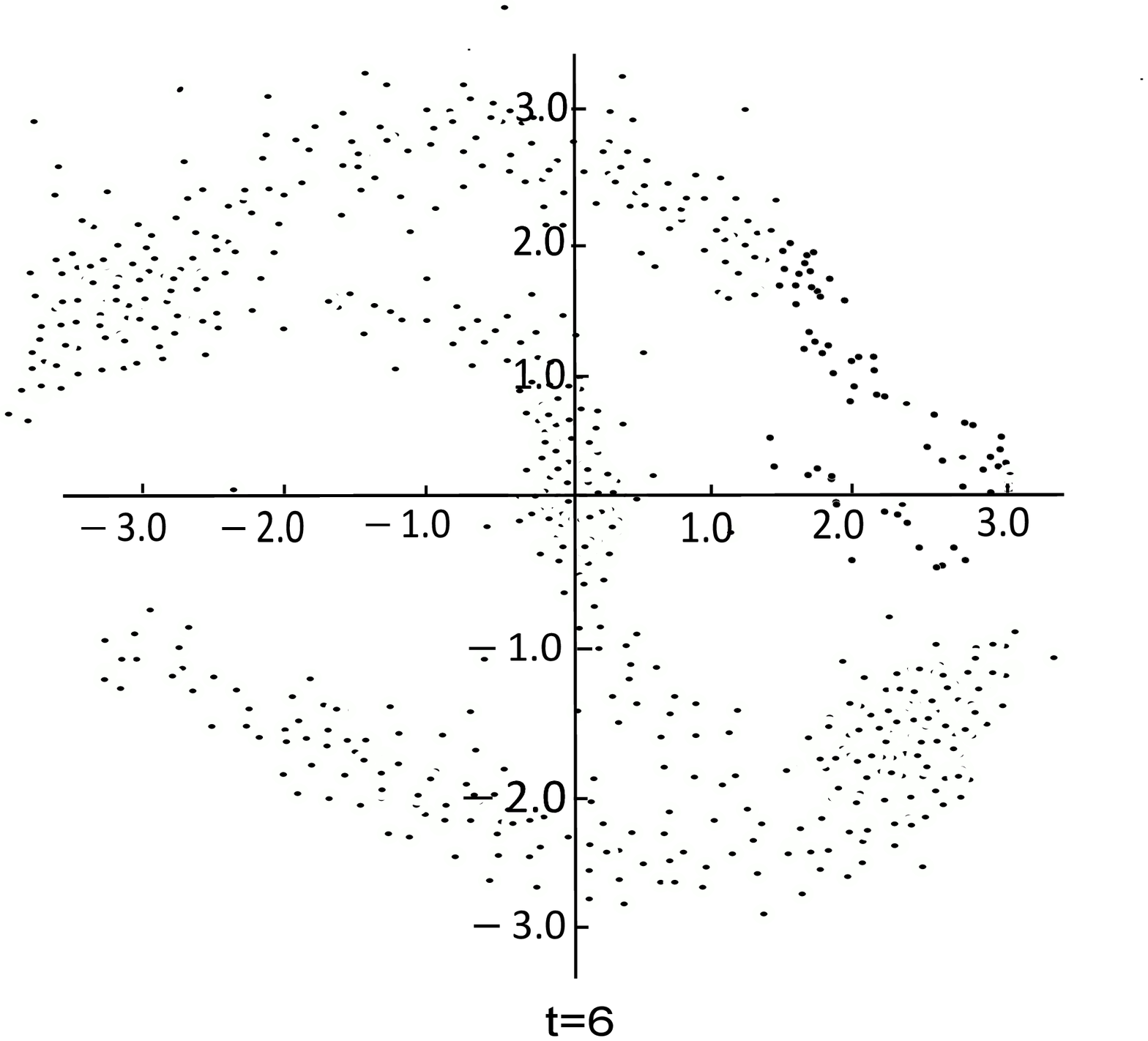}

Figure 1 Formation of the spiral substructure
\end{center}

\begin{tabular}{cc}
\includegraphics[height=200pt, width=180pt]{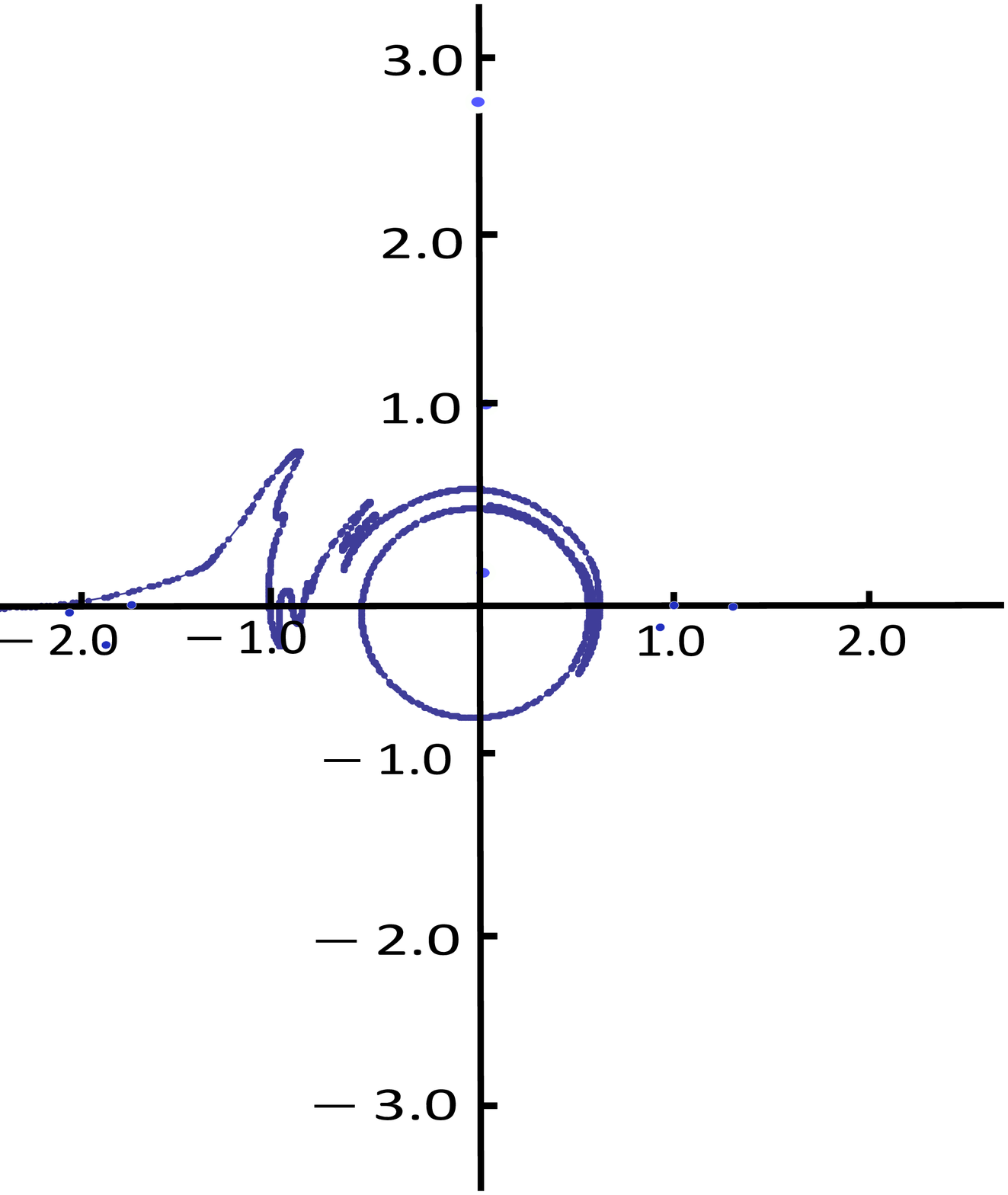} &
\includegraphics[height=200pt, width=180pt]{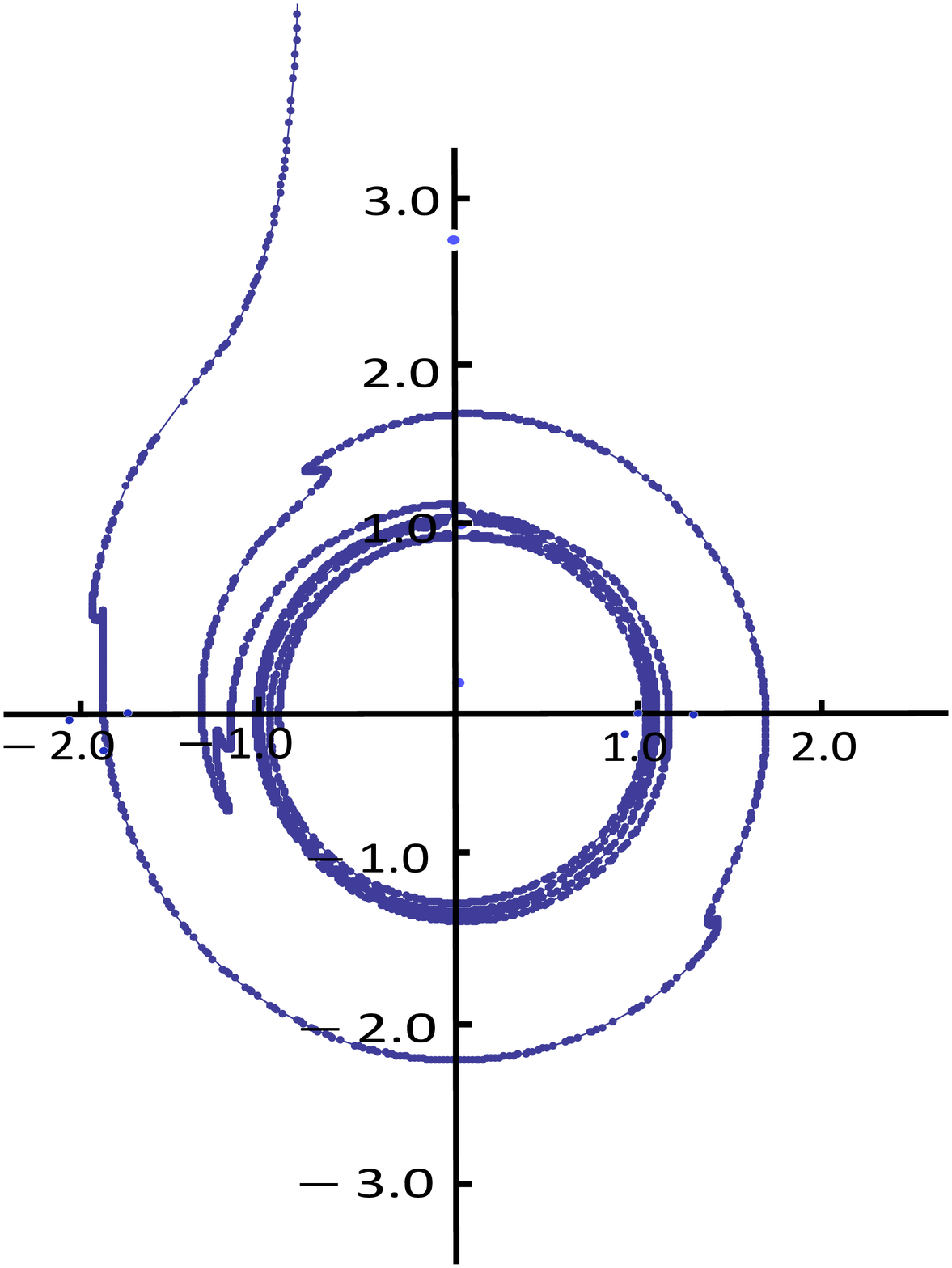} \\
\end{tabular}

\begin{center}
Figure 2
\end{center}

The axisymmetric disk on which a spiral pattern is superimposed
then stabilizes; the system becomes quasi-steady after approximately ${2\
10}^9$ years ($t=20$). An impressive large-scale coherent "grand design" galaxy
appears. This is evidenced in figures 3.b and 3.c.

\begin{center}
\includegraphics[height=200pt, width=400pt]{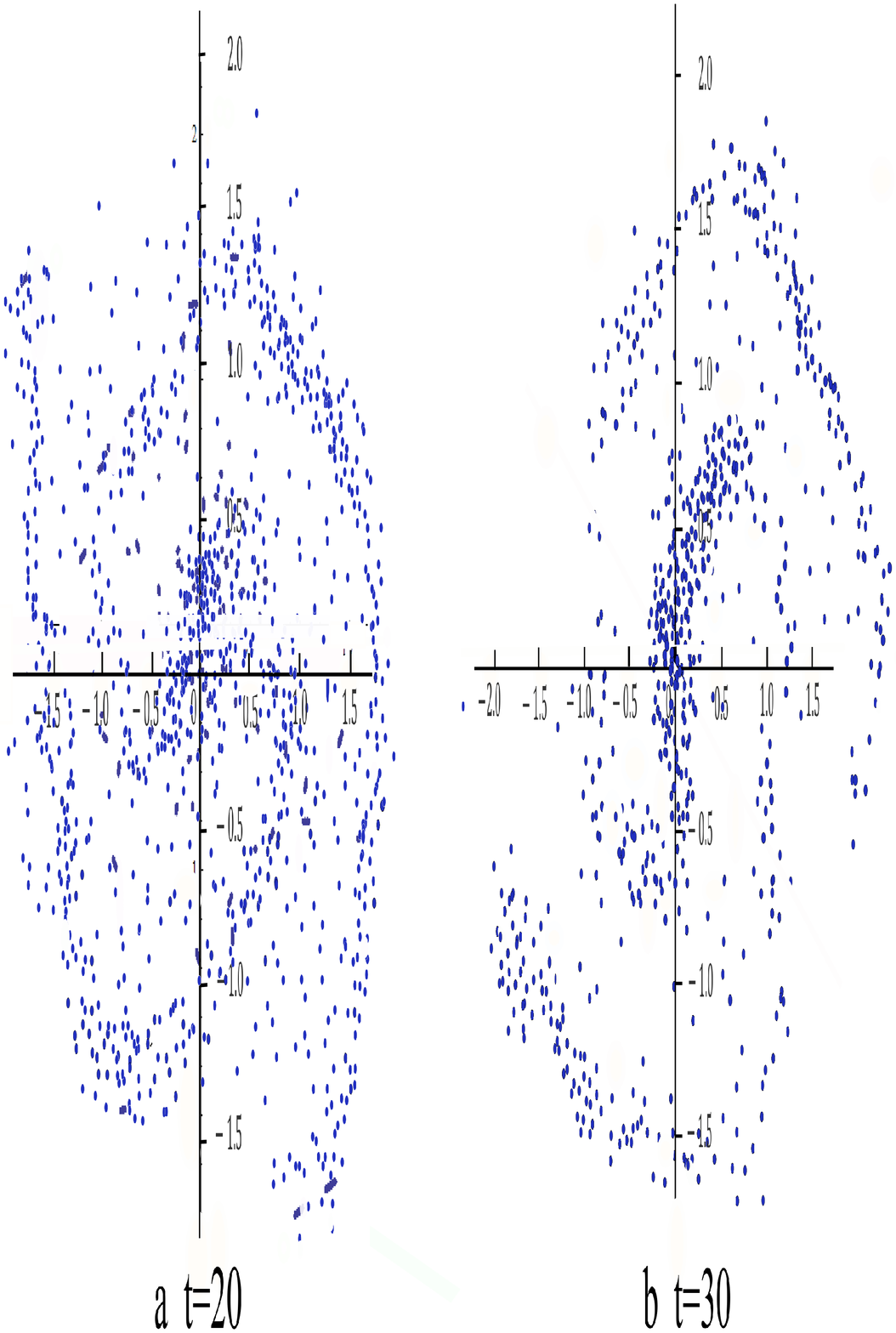}

\vspace{10pt}
Figure 3 The quasi-permanent spiral in the \texorpdfstring{$\kappa$-model}{Lg}
\end{center}

 That the spiral pattern and
the central bulge are self-maintained with time is also clearly seen in
these figures. This result is quite remarkable, and this is not the case within the
framework of the dark matter paradigm, where the spiral substructures are just
transient phenomena over a few rotational periods of the galaxy. However, some local and relatively strong deformations, such as
those due to self-gravity, persist even after two milliards of
years, and the galaxy after formation does not rotate as a rigid body,
in contrast to what one might think in the framework of the $\kappa$-model. There
remain a very large number of stars that exhibit elliptic, not fully
circularized, motions, a phenomenon that also contributes to the deformations. In the
framework of the $\kappa$-model, a galaxy is a "living" object, and it should
not be seen as a perfect and immutable wheel. Accordingly, the substructure
spiral is always evolving, albeit much more slowly than in other contexts, such as,
for instance, in density wave theory, where the lifetime of the spiral
substructure appears very short unless the density wave is constantly
fed.

In parallel, a quasi-flat rotation curve for the velocities is obtained (figure 4; $r={\kappa{}}_E\sigma{}$). Let us note that for practical
(observational) reasons, we do not report the velocities in this figure but rather
the measured shift of the frequencies $\nu{}$. Let us specify that the
radial velocity measurements supply the true velocities (in fact, the radial part),
i.e., those measured by an inertial observer on site\footnote{The curve that is displayed is not properly covered for $r \leq 0.3$. A three
dimensional model for the bulge is needed. Curve(2) is obtained from the mass density derived by resolving
equations (1) and (2) and then recalculating the velocities.}. We have also reported Keplerian velocity curve (2) in the same figure, which makes $\kappa{}=\kappa_E$ everywhere. However, for $r\gtrsim 5,\ \kappa{}$ becomes constant again, and rotation
curve (1) falls off in a Keplerian manner (not shown in the figure).

\vspace{3pt} \noindent
\begin{tabular}{p{221pt}p{210pt}}
\parbox{212pt}{\centering
\hspace{-30pt} \includegraphics[height=144pt, width=215pt]{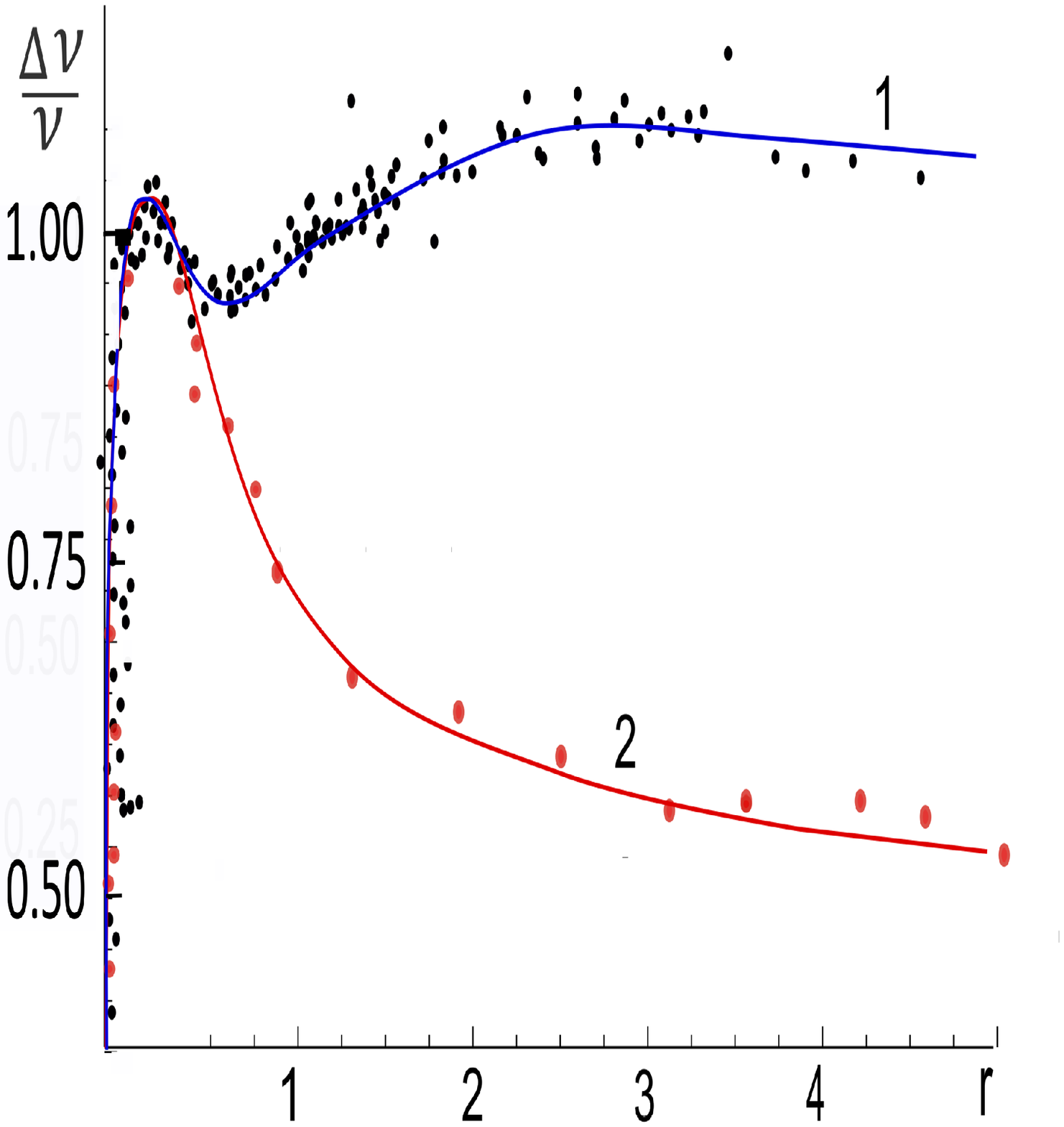}

{\vspace{-30pt}
{Figure 4 Galactic rotation curve:}
}} & \parbox{210pt}{

\includegraphics[height=90pt, width=145pt]{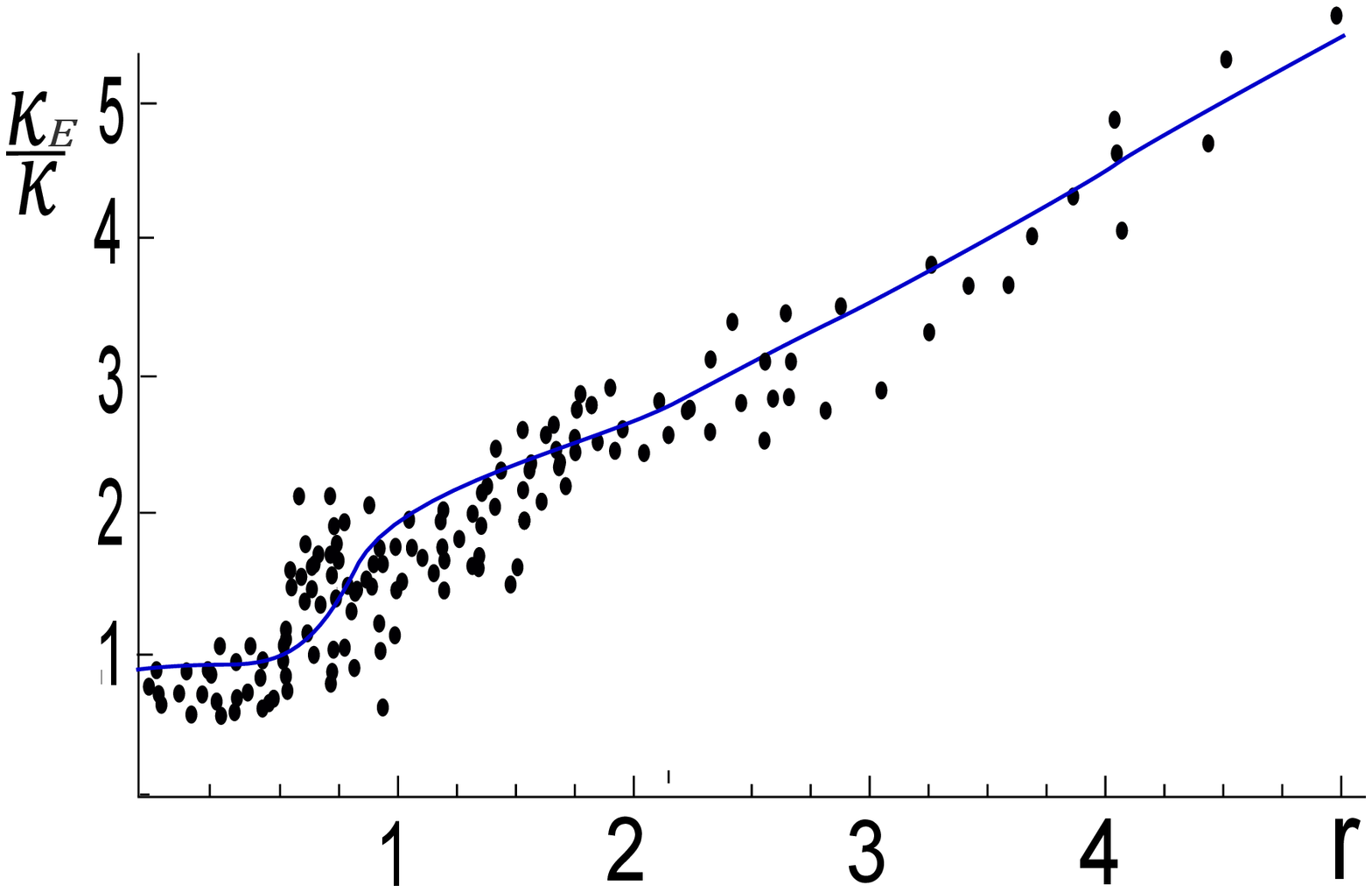}
\vspace{-10pt}

} \\
\parbox{210pt}{{(1) $\kappa$-model \ \ (2) pure Newtonian model}

} & \parbox{210pt}{\vspace{-10pt} Figure 5 \\ {Variation in $\kappa_E/\kappa{}$ along a galactic radius}
} \\
\end{tabular}

\subsection{ The resolution of the winding dilemma}

The winding problem has always been recurrent in the simulation of "grand design" galaxies (Shu, 2016) and remains partially unsolved. However, against all odds, the two interlinked effects, i.e., the flatness of the
rotation curve and the almost steady nature of the spiral design, can easily be understood in the framework of the $\kappa$-model. Let us define two inertial observers $A$ and $B$
situated along a galactic radius, each of them measuring the
(true) velocity of one particle, resp. $M$ for $A$ and ${M}'$ for $B$, located near them. We have

\[
v_A\left(M\right)=v_B\left(M'\right)
\]

{\raggedright expressing the flatness of the rotation curve as measured by
spectroscopy}\footnote{In the Universe, the radial velocities are estimated from
spectroscopic measurements. These velocities are the same as those measured by a
local inertial observer. However, the tangential velocities are
deduced from proper motions and parallax measurements. These apparent
quantities are linked to the terrestrial observer who measures them by
trigonometry, postulating a unique, and most likely imaginary, background.
These two methods are very different, and this difference is expressed in the
framework of the $\kappa$-model. The first method (spectroscopy) supplies true
(radial) velocities, whereas the second method (observations of proper motions)
supplies apparent (tangential) velocities.}).

However, $v_A\left(M\right)={\kappa{}}_A\sigma{}(M)\dot{\theta{}}(M)$ and $v_B\left(M'\right)={\kappa{}}_B\sigma{}(M^{'})\dot{\theta{}}(M')$, where $\theta{}$ is
the polar angle defined from a radial baseline with origin at the
galaxy center (in the $\kappa$-model, a given direction is well identified and is
the same for all observers). Following figure 3, the coefficient $\kappa{}$ (or more rigorously, the measurable ratio $\frac{\kappa}{\kappa_E}$)
is approximately proportional to $\frac{1}{\kappa_E\sigma{}}$ in the outer
regions of a galaxy\footnote{More rigorously, this coefficient is a mean $\kappa{}$ obtained
by curve fitting (the curve in figure 5). }; thus, we obtain

\begin{equation}
\dot{\theta{}}(M)=\dot{\theta{}}(M')
\end{equation}

{\raggedright for any pair of points located along a galactic radius. The points remain
steadily aligned with time along the galactic radius (a straight line plotted
from the galactic center). This noteworthy outcome enables us to now
understand why we can simultaneously observe a quasi-steady "grand
design" substructure and a flat rotation curve, two effects that seemingly
to contradict each other. Let us note, however, that the
persistence of the spiral substructure is not absolute and that ultimately, the
latter may slowly distort with time. The main reason for this is that the orbits
of stars are not perfect circles but rather ellipses with slight eccentricities.
On the other hand, the factor $\kappa{}$ fluctuates with the
variation in the density within the spiral substructure.}

In this respect, comparison with the results obtained in the framework of
the dark matter paradigm is quite interesting. In the latter simulations, we 
obviously take the same initial conditions for the baryonic matter. We can
conclude from this comparison that the $\kappa$-model yields something similar to
the results from the dark matter paradigm in many aspects but without dark matter; however
the spiral substructure is self-maintained and is much more impressive.

To perform this comparison, we add a halo of dark matter with a
density distribution proportional to $r^{-2}$ in the outer regions (halo mass)
and we take $M_{DM}\sim{}10\ M_B$ and $\kappa{}=\kappa_E{}$ everywhere, i.e., the
orthodox strategy for rehabilitating the common background, where we can set $r = \kappa_E\sigma{}$.
Then, with the dark matter superimposed on the baryonic
component taken into account, we can see that a spiral pattern with a central
bar effectively appears to be essentially due to self-gravity, but there are now are now multiple, transient and discontinuous outer arms with very short lifetimes (figures 6.a, 6.b).

\begin{center}
\includegraphics[height=200pt, width=350pt]{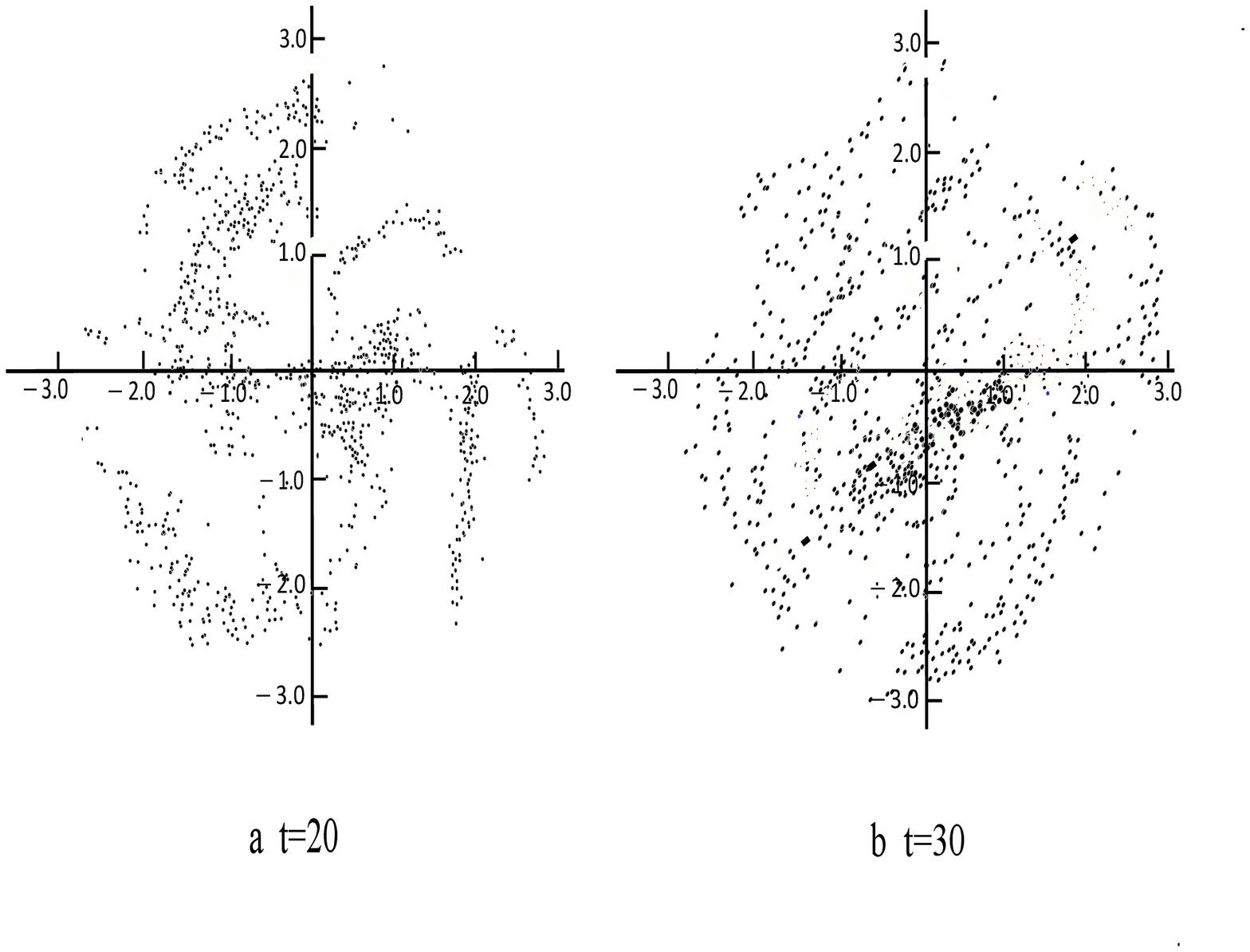}
\end{center}

 \begin{center}

{Figure 6      The transient spiral in the dark matter model}
\end{center}

In any case, a flocculent galaxy pattern seems to emerge from the
simulations, but definitely not a permanent "grand design". Certainly, at $t=20$,
figure 6a clearly mimics a spiral-type structure but with four arms
instead of two; however, at $t=30$, the situation becomes confusing; the
galaxy core is getting denser, and the arms are more spiraled and  are unrecognizable.
This effect is well known, and the same conclusion is drawn by the other
published N-body simulations within spherical dark matter models, even
those using very sophisticated SPH, grid-based procedures, including
complicated physics such as cooling, star formation, and energy injection from
supernovae (Guedes, Callegari, Madau and Mayer, 2011). We can compare figure 6 of the
present paper and figure 2 of (Guedes, Callegari, Madau and Mayer, 2011) obtained after a series
of state-of-the-art simulations; on the one hand, we see that they are fairly
similar, but on the other hand, we are far from the set objective
for a realistic "grand design" galaxy. Obviously, additional aspects
can be artificially introduced in the calculations  such as a permanent
density wave within the gas driven by a rotating central bar (Shu, 2016), a
triaxial distribution of dark matter forming a rotating halo
with angular momentum (Martinez-Medina, Bray and Matosa, 2015), or a Yukawian gravitational
potential (Brandao and de Araujo, 2012). However, even though all these proposals are
interesting, a solid experimental/observational basis is still lacking. The
interaction with a galaxy companion to sustain a "grand design"
substructure may be much more realistic (Dobbs, Theis, Pringle. and Bate, 2010), even though there
exist cases of "grand design" galaxies without the presence of a
companion.

By contrast, starting from an initial homogeneous cold gas pervaded with a
shear velocity field, it is remarkable that, in a very simple way, the
$\kappa$-model naturally directly leads to a quasi-steady "grand design" spiral
galaxy with very clearly well-formed and distinct arms. A bar also appears in
the central region, and the arms are formed by self-gravity, which is
much higher in the outer regions (weak mean density) than in the inner regions
(high mean density) of the galaxy. Another noticeable point is that the spiral
substructure becomes stronger with time instead of the weakening phenomenon
generated in other models with dark matter or even in MOND or in MOG. These
results provide strong support for the self-consistent $\kappa$-model, the
important conclusion being that the knowledge of the sole distribution of
visible baryonic matter is sufficient to understand the dynamics of a galaxy and
that the introduction of exotic particles is not needed.

\section{Two selected examples}

The preceding simulation can, however, appear to be a toy model. Most interesting is the
practical application. For this, we choose two examples, a high luminosity galaxy
(New General Catalogue of Nebulae and Clusters of Stars (NGC) 6946) and a low luminosity galaxy (NGC 1560). These two examples are taken
from the review of Famaey and McGaugh (2012, see the references therein for the original data sources).

\begin{center}
\includegraphics[height=200pt, width=200pt]{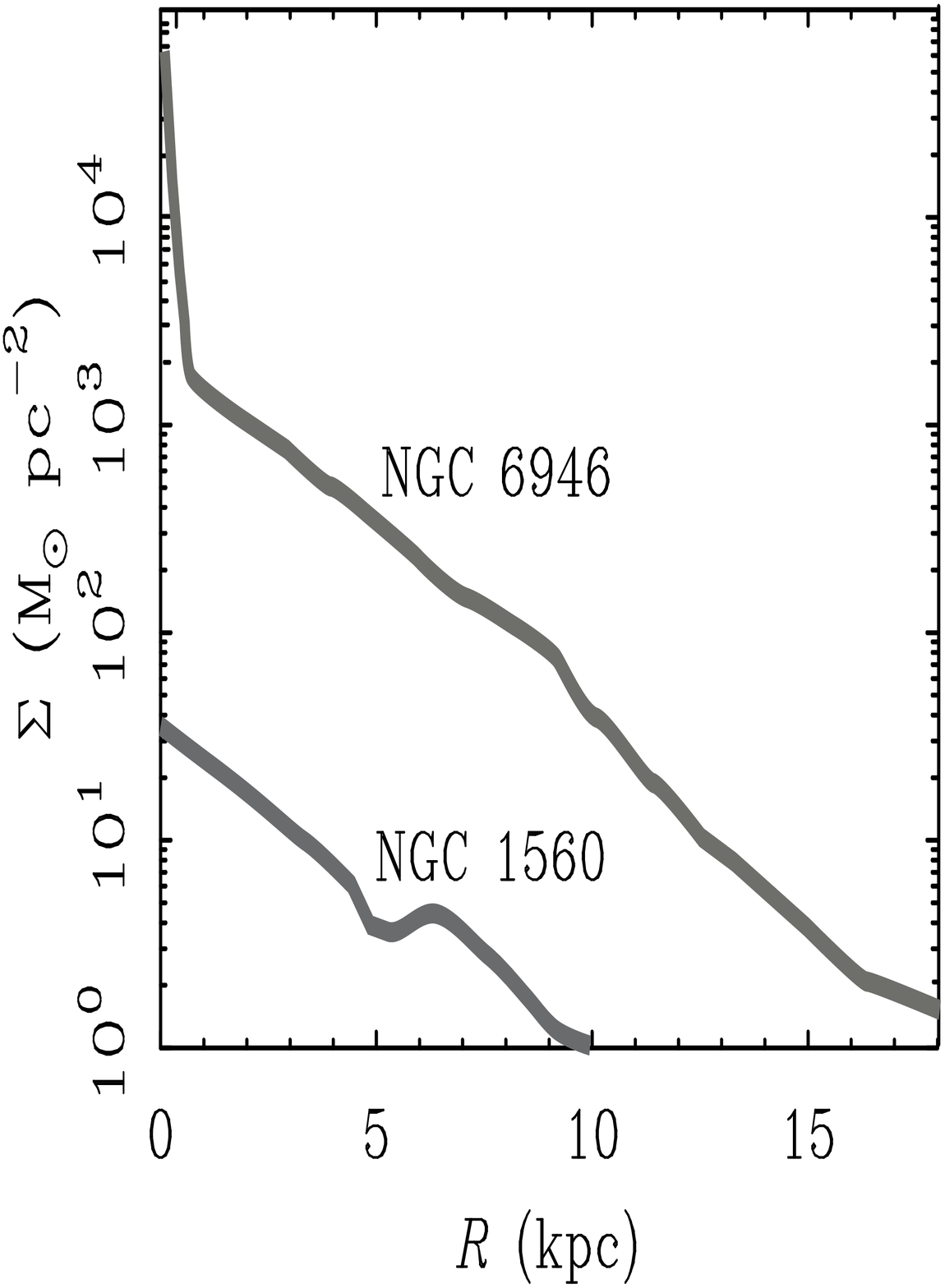}

\end{center}

Figure 7 Surface density profiles (stars + gas) of two galaxies: the high surface brightness (HSB)
spiral NGC 6946 and the low  surface brightness (LSB) galaxy NGC 1560 (from Famaey and McGaugh, 2012, figure 13)

\vspace{10pt}

We assume here that the galaxies to be studied are stabilized and that the orbits
of the stars are circular. The Newtonian velocities $v_{Newt}$ could then be obtained from equation (1), making $\kappa_i\equiv \kappa_E$, , $\forall i$ in it\footnote{We can see that in equation (1), the factor $(\frac{\kappa_E}{\kappa_i})$ is factorizable in front of the sum over $j$. This sum expresses the Newtonian force acting on a particle with index $i$.}. The true velocities $v$ (measured by spectroscopy)
are then deduced from the Newtonian velocities $v_{Newt}$ by the following:
"magnification" relation (this deduction is made in appendix B)

\begin{equation}
v={\left(\frac{{\kappa{}}}{\kappa{}_E}\right)} {\left(\frac{{\kappa{}}_E}{\kappa{}}\right)}^{\frac{3}{2}}v_{Newt}={\left(\frac{{\kappa{}}_E}{{\kappa{}}_0}\right)}^{\frac{1}{2}}{\left(\frac{{\kappa{}}_0}{\kappa{}}\right)}^{\frac{1}{2}}v_{Newt}
\end{equation}

{\raggedright where}

\begin{equation}
\frac{{\kappa{}}_0}{\kappa{}}=1+Ln(\ \frac{{\Sigma{}}_0}{\Sigma{}}\
\frac{\delta{}}{{\delta{}}_0})
\end{equation}

Equation (7) directly derives from equation (2), where $\Sigma{}$ represents the surface density and
$\delta{}$ the mean thickness of the matter (stars + gas) along the line of
sight. The index $0$ represents the maximum value of $\Sigma{}$ for a given
galaxy. Assuming a uniform thickness, we take
$\delta{}={\delta{}}_0=Const$ throughout the calculations. Figure 8 displays
the ratio $(\frac{{\kappa{}}_0}{\kappa{}})^{\frac{1}{2}}$ for the two selected galaxies calculated
with equation (7).

\begin{center}
\includegraphics[height=200pt, width=200pt]{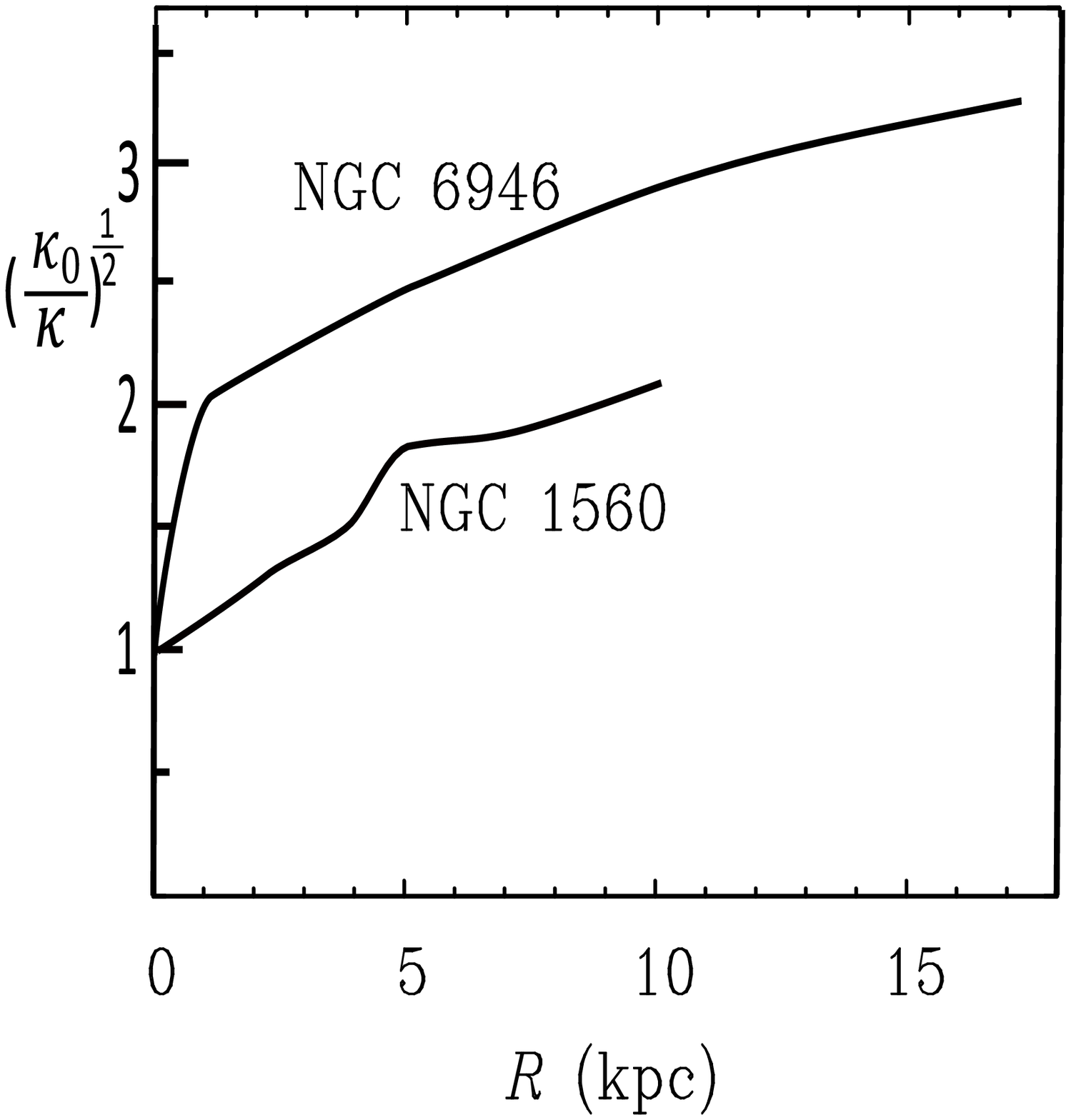}

\end{center}

Figure 8 Variation in $(\frac{{\kappa{}}_0}{\kappa{}})^{0.5} $ as a function of the
radial distance  $R$

\vspace{10pt}

The global ratios $\frac{{\kappa{}}_E}{{\kappa{}}_0}$ are unknown, but we can attempt to supply an estimate by again employing equation (2)\footnote{The relation that has then been used is

\[
\frac{{\kappa{}}_1}{{\kappa{}}_2}=1+Ln(\ \frac{{\Sigma{}}_1}{{\Sigma{}}_2}\
\frac{{\delta{}}_2}{{\delta{}}_1})
\]

respecting the condition
${\Sigma{}}_1{\delta{}}_2>{\Sigma{}}_2{\delta{}}_1$.}. With   ${\Sigma{}}_{\odot}=60
\ M_{\odot}\ {pc}^{-2}$ (from Famaey, and McGaugh, 2012, figure 19; Moni Bidin et al, 2012, figure 5), we find that
${\left(\frac{{\kappa{}}_E}{{\kappa{}}_0}\right)}^{0.5}=0.36$    for  NGC 6946
and $1.24$ for NGC 1560. We, however, note that a better fit of the
$v$-curves is realized by taking
${\left(\frac{{\kappa{}}_E}{{\kappa{}}_0}\right)}^{0.5}=0.46$ (instead of
$0.36$) for NGC 6946 and $1.30$ (instead of $1.24$) for NGC 1560. In any case, the calculation of these global magnifications
factors is certainly vitiated by various biases.

 For instance, we know that the inclination
angle $i$ of a galaxy can strongly impact the velocity curves $\left[26\right]$. For NGC 6946, a variation of 
$7^\circ{}$ in the inclination angle, starting from $i=38^\circ{}$ for this galaxy $\left[26\right]$,
is equivalent to increasing the factor ${\left(\frac{{\kappa{}}_E}{{\kappa{}}_0}\right)}^{0.5}$ from the value predicted by the $\kappa$ model ($0.36$) to its fit value ($0.46$). Eventually, even though to a lesser extent, the distance of this galaxy is highly uncertain (Elridge and Xiao, 2019). This important parameter intervenes in the estimate of the absolute luminosity, subsequently in that of the surface density, and eventually in that of the ratio ${\left(\frac{{\kappa{}}_E}{{\kappa{}}_0}\right)}^{0.5}$.

The estimate
of ${\Sigma{}}_{\odot}$ can also play a role. Thus, for NGC 1560, passing from
${\Sigma{}}_{\odot}=60$ to $70\  M_{\odot}\ {pc}^{-2}$ increases the numerical value for
${\left(\frac{{\kappa{}}_E}{{\kappa{}}_0}\right)}^{0.5}$ from $1.24$    to the value taken for the fit in figure 10, $1.30$ (with an insignificant change for NGC 6946). Note that a value $\sim 90 \  M_{\odot}\ {pc}^{-2}$ is given in [Sofue, Honma and Omodaka, 2009, Table 4]. However, this value is well above the estimates of other authors (Famaey and McGaugh, 2012; Moni Bidin et al, 2012) and is very likely overestimated.

Fortunately, while some uncertainty is associated with the global ratio ${\left(\frac{{\kappa{}}_E}{{\kappa{}}_0}\right)}^{0.5}$, the resulting effect is a global shift of the mean height
of the velocity curve\footnote{In the literature, the mean height of the observational velocity curve can also vary substantially; for instance, for the Milky Way, we can compare (Sofue, Honm and Omodaka, 2009, figure 1) and (McGaugh, 2016, figures 5,6]}, which does not heavily affect the details of
the curve, such as bumps, wiggles, or the typical flatness in the outer regions. The final results are presented in figure 9 (NGC 6946) and figure  10 (NGC 1560).

\begin{center}
\includegraphics[height=200pt, width=200pt]{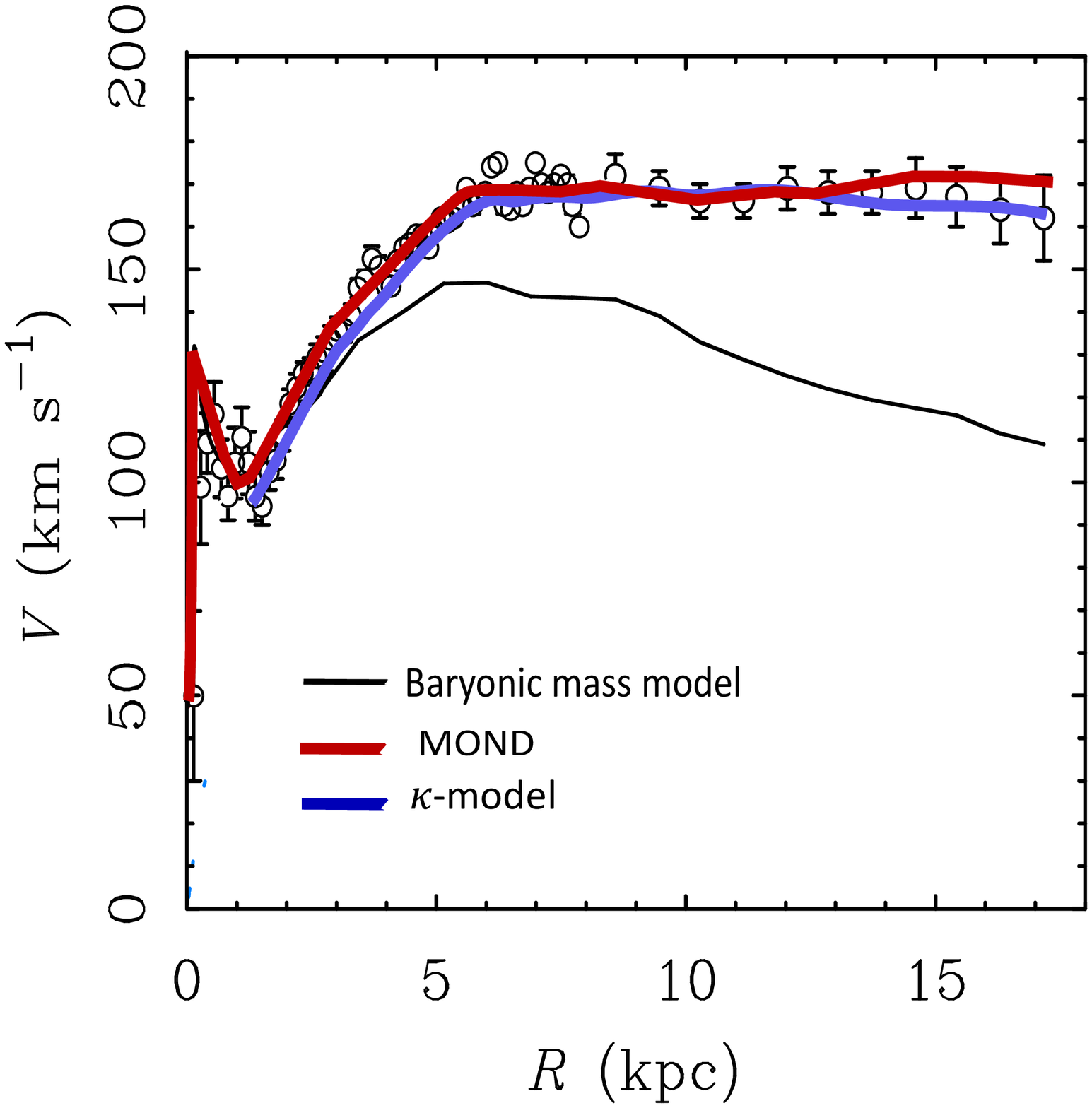}

\end{center}

Figure 9 NGC 6946: the observed rotation curve (black
circles) is displayed together with those predicted by the
baryonic mass model (black line), MOND (red line) and $\kappa{}$-model (blue line)

\vspace{10pt}

By examining the curves reproduced in figure 9, we can see that MOND and the
$\kappa{}$-model give very similar values for the velocities between
$6>r>13\ kpc.$  For $r<6\ kpc$, MOND is slightly better than the
$\kappa{}$-model, in contrast to $r>13\ kpc$, where the MOND curve does not
initiate a slow decrease, while this decrease is very
noticeable in the curve of the $\kappa{}$-model. However, the differences
between MOND and the $\kappa{}$-model are small.

Regarding galaxies of lower baryonic surface densities, it is well known that Newtonian prediction is very poor everywhere, even at smaller radii [29]. The galaxy NGC 1560 illustrates this situation,
whereas both MOND and the $\kappa{}$-model are fairly consistent with the observed
curve (figure 10).

\begin{center}
\includegraphics[height=200pt, width=200pt]{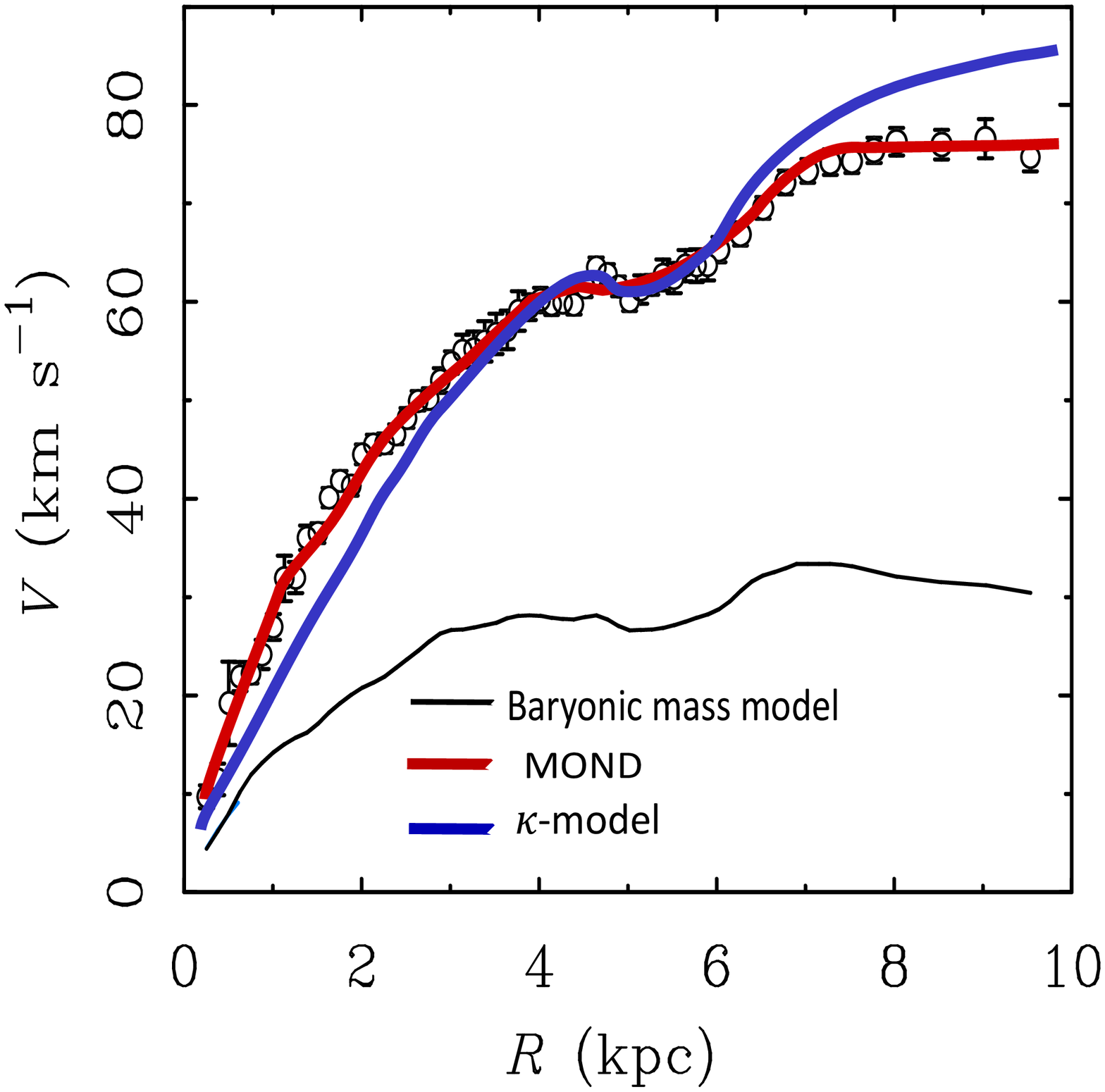}

\end{center}

Figure 10 NGC 1560: the observed rotation curve (black
circles) is displayed together with the theoretical curves predicted by the baryonic mass model (black line), MOND
(red line) and $\kappa{}$-model (blue line)

\vspace{10pt}

The results from both MOND and the $\kappa{}$-model are 
superimposed well between $3>r>6\ kpc$, even though outside this interval,
MOND seems slightly better. However, an optimal overlap of the
$\kappa{}$-model versus the observation could eventually be realized by an
adjustment of the thickness $\delta{}$. Thus, the $\kappa{}$-model could predict that
the thickness along the line of sight, $\delta{},$ is larger than
${\delta{}}_0$ for $1>r>3\ kpc$ by a factor $1.5$ and conversely smaller by the same factor for $6>r>10\ kpc$.

 However, let us note that a comparison with the observational curve derived from other sources (S\'{a}nchez-Salcedo and Hidalgo-G\'{a}mez, 1999) can lead to another interpretation of the results. The x-axis of figure 10 has been adjusted for figure 11.  In this case, we can see that within the intervals $0>r>3\ kpc$ and $8>r>10\ kpc$, the observational data are now located in the area bounded by MOND and the $\kappa$ model, namely, resp. above by MOND and below by the $\kappa$-model for $0>r>3\ kpc$ and the opposite for $6>r>10\ kpc$.

\begin{center}
\includegraphics[height=260pt, width=210pt]{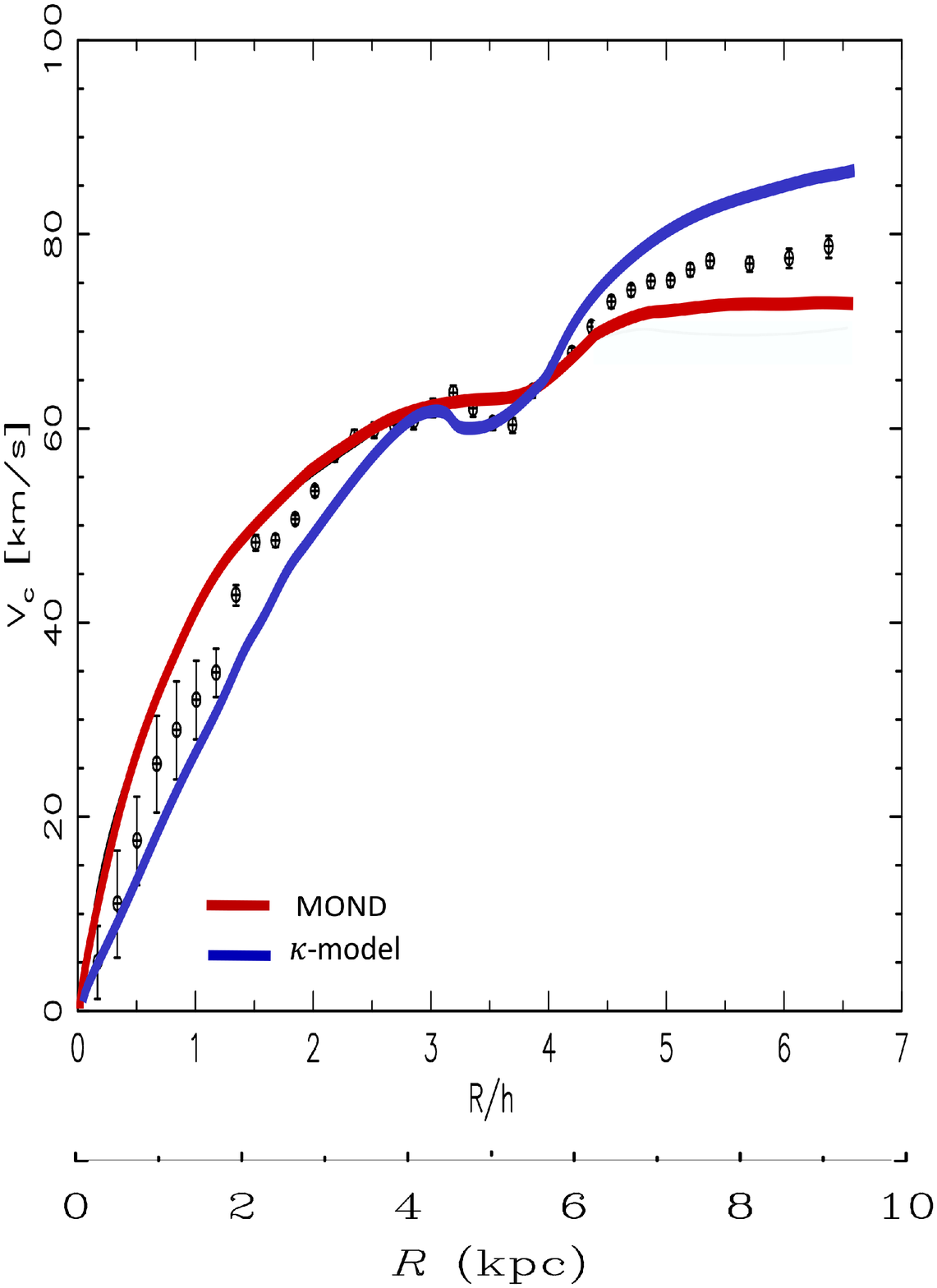}

\end{center}

Figure 11 NGC 1560: Another observed rotation curve (black
circles), derived from (S\'{a}nchez-Salcedo and Hidalgo-G\'{a}mez, 1999), is displayed together with the theoretical curves predicted by MOND (red line, reproduced from S\'{a}nchez-Salcedo and Hidalgo-G\'{a}mez (1999)) and the $\kappa{}$-model (blue line)

\vspace{10pt}

\section{AGC 114905: an ultradiffuse galaxy without dark matter?}

The ultradiffuse galaxies are LSB galaxies with an
extended light distribution (Conselice, 2018; Mancera Piña et al, 2021). Belonging to this peculiar category of galaxies, AGC 114905 seems to pose a challenge to the standard dark matter model as well as to MOND. Figures 12 and 13 show the high-resolution HI interferometric observations obtained by Mancera Piña et al (2021) for AGC 114905.

\begin{center}
\includegraphics[height=200pt, width=200pt]{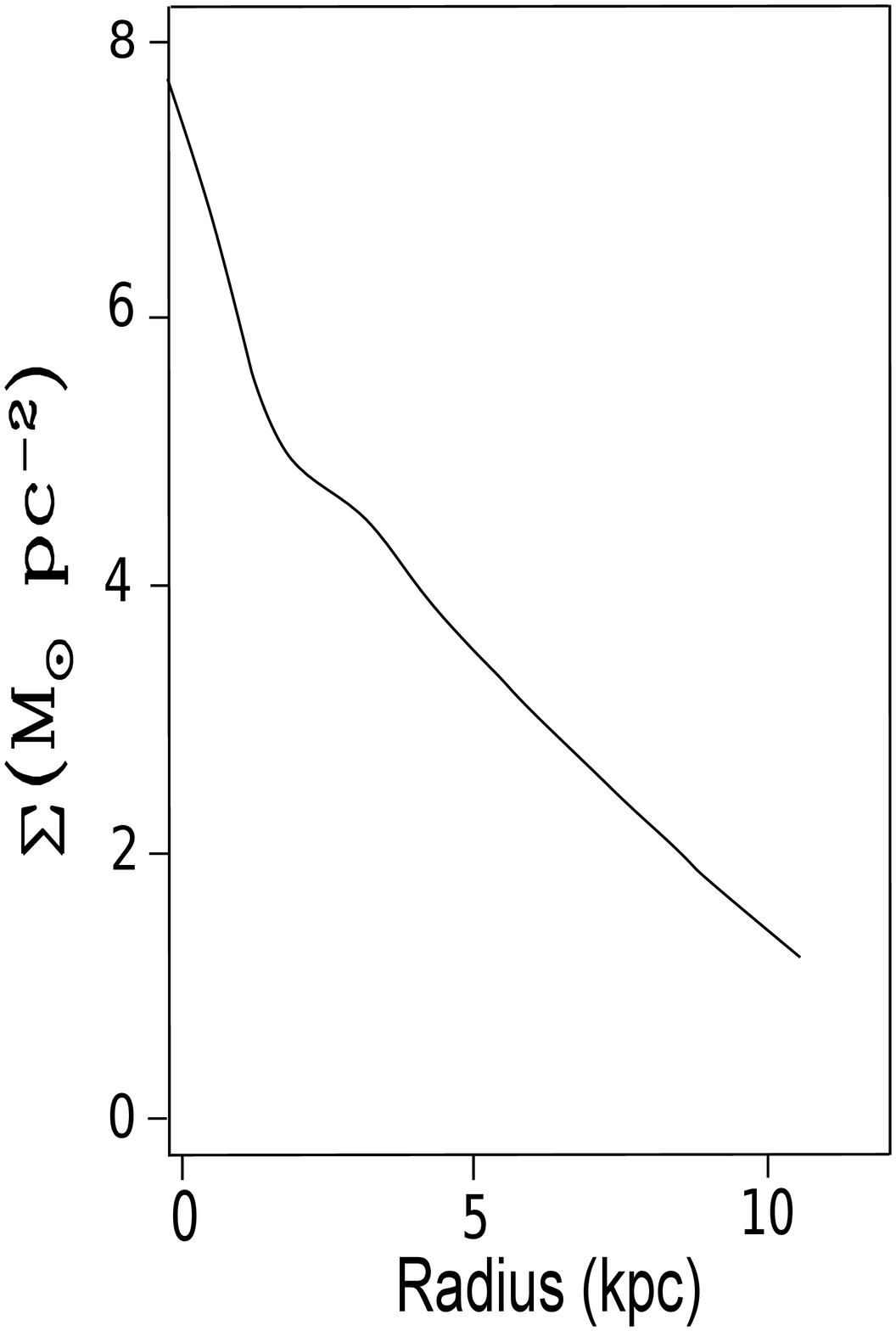}
\end{center}

Figure 12 SMD profile (stars + gas) of AGC 114905 (from figure 1 of Mancera Piña et al, 2021).

\begin{center}
\includegraphics[height=200pt, width=200pt]{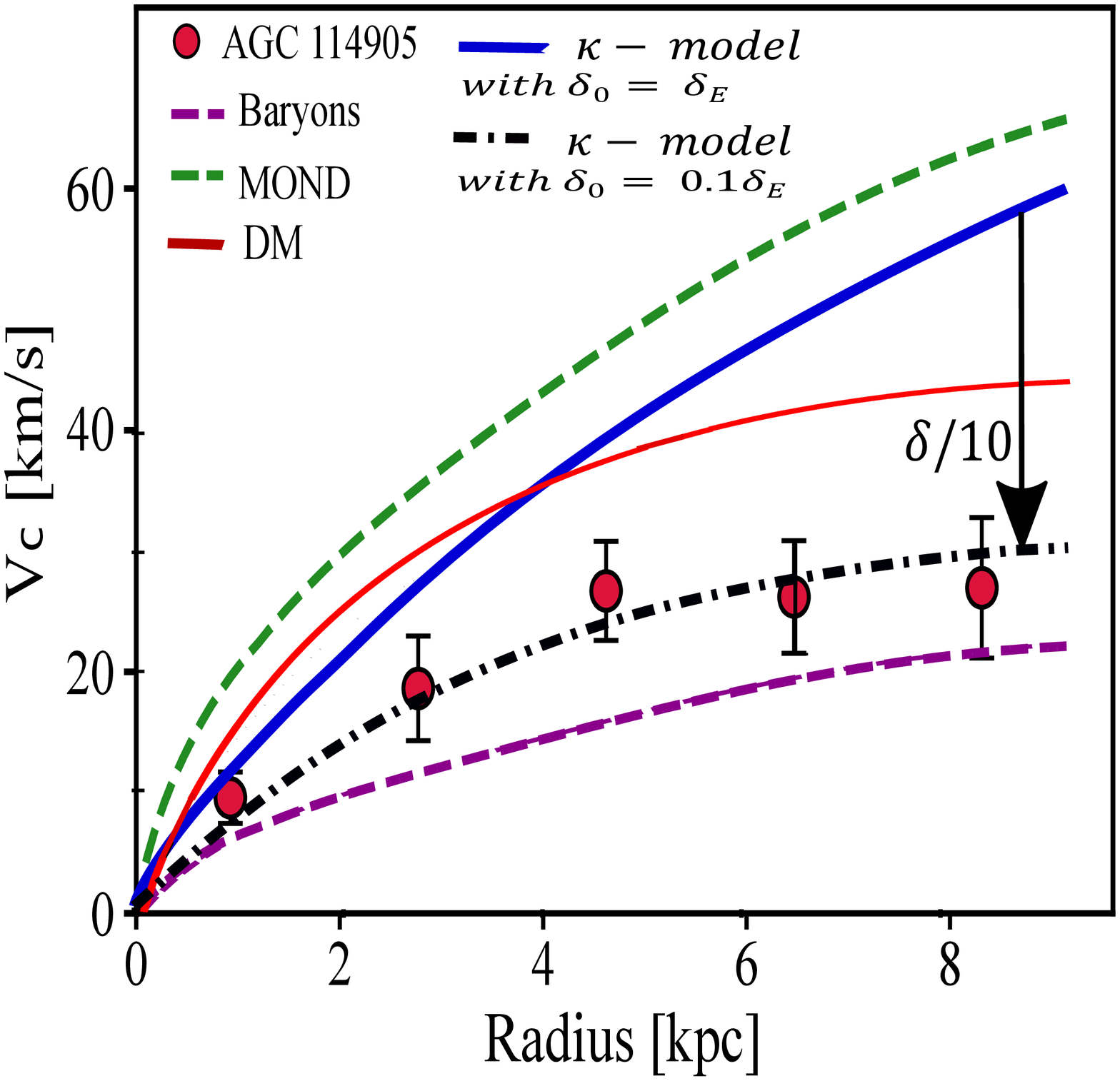}
\end{center}

Figure 13 The rotation curve of AGC 114905 (from figure 5 of Mancera Piña et al, 2021). The red points show the observational data. The inclination angle is equal to $26^\circ$. The dark matter halo is represented by a red line, MOND by a green dashed line and the $\kappa$-model by a blue line.

\vspace{10pt}
As shown in figure 13, we note that MOND and the $\kappa$-model give very similar profiles. The agreement between MOND and the $\kappa$-model, already identified in the preceding section, naturally underlines that the domain of weak accelerations (MOND) is related to the domain of weak mean densities (the $\kappa$-model).

Nevertheless, we see that all three theoretical models, dark matter, MOND and $\kappa$, fail to represent the observational curve. Mancera Piña et al (2019) suggest that a drastic modification of the inclination angle, from $26^\circ$ within the range $10^\circ -15^\circ$, would be needed in order
to find agreement between the dark matter or MOND prediction and the observational profile
of AGC 114905. The same reasoning can be applied to the $\kappa$-model. These authors further stated that a radially varying inclination could potentially alleviate the
tension between the rotation curve shapes, especially for MOND. However, Mancera Piña et al, (2021) also specify that inclinations as low
as $10^\circ - 15^\circ$ are very likely inconsistent with the observational data.

Thus, even though the discrepancy between the $\kappa$-model and the observational data could be reduced by lowering the inclination angle, Mancera Piña et al (2021) find this proposal unreasonable. However, apart from the inclination parameter, the thickness of the galaxy plays a role in the $\kappa$-model because the latter factor influences the global magnification ratio ${\left(\frac{{\kappa{}}_E}{{\kappa{}}_0}\right)}^{0.5}$ (equation (7)). For instance, a drastic reduction in the mean thickness $\delta$ by a factor $10$ (at constant $\Sigma$  given this quantity is measured) would bring the $\kappa$-model curve within the error bars of the observational data (figure 13). Thus, rather than seeing   AGC 114905 as a galaxy deprived of dark matter, such an explanation, i.e., by considering the mean thickness, would lead to envisaging this galaxy as a very smeared and hence ultrathin object.

\section{The Milky Way}

The curve of rotation of the Milky Way has been extensively studied, but the
observational curves appear to strongly vary with the type of disk
tracer. We can compare, for instance, figure 2 of
Bhattacharjee, Chaudhury and Kundu (2014),    figure  11
of Huang et al. (2016) or figure  1(d) of Sofue (2020), where we can see that the data exhibit
extensive scattering. Here,
our aim is to fit the panel of observational rotation curves of the Milky Way with the
$\kappa{}-$model. For that, the baryonic surface mass density (SMD) is needed.
This is the sole component given that we exclude the dark matter halo. We
use two $SMD$ profiles, one from Famaey and Mc Gaugh
(2012$)$ and the other very close to the
$SMD$ profile of Sofue (2020, figure 2).

\begin{itemize}
	\item \textbf{The SMD for the distances $8<r<12\ kpc$}
\end{itemize}

Figure 14 reproduces the SMD of Famaey and Mc Gaugh (2012, figure 29). Other
profiles are also supplied in Mc Gaugh (2016), but these profiles are mostly the
same except for a few details. The results are displayed in figure 15. We adopted a global magnification factor of $0.46$ (against $0.43$ for NGC 6946). We observe that both MOND and the $\kappa{}-$model
adequately reproduce the sequence of wiggles of observational curve.
These wiggles, which are already present in the Newtonian curve, are found
in the observational curve following Sencisi's rule (Sancisi, 2004). In the
framework of the $\kappa{}-$model, the interpretation of this effect is that
the observational curve is simply a magnification of the Newtonian curve, a
phenomenon that is very well illustrated here. Let us note that this patent
fact remains unexplained in the framework of dark matter models.

\begin{center}
\includegraphics[height=200pt, width=200pt]{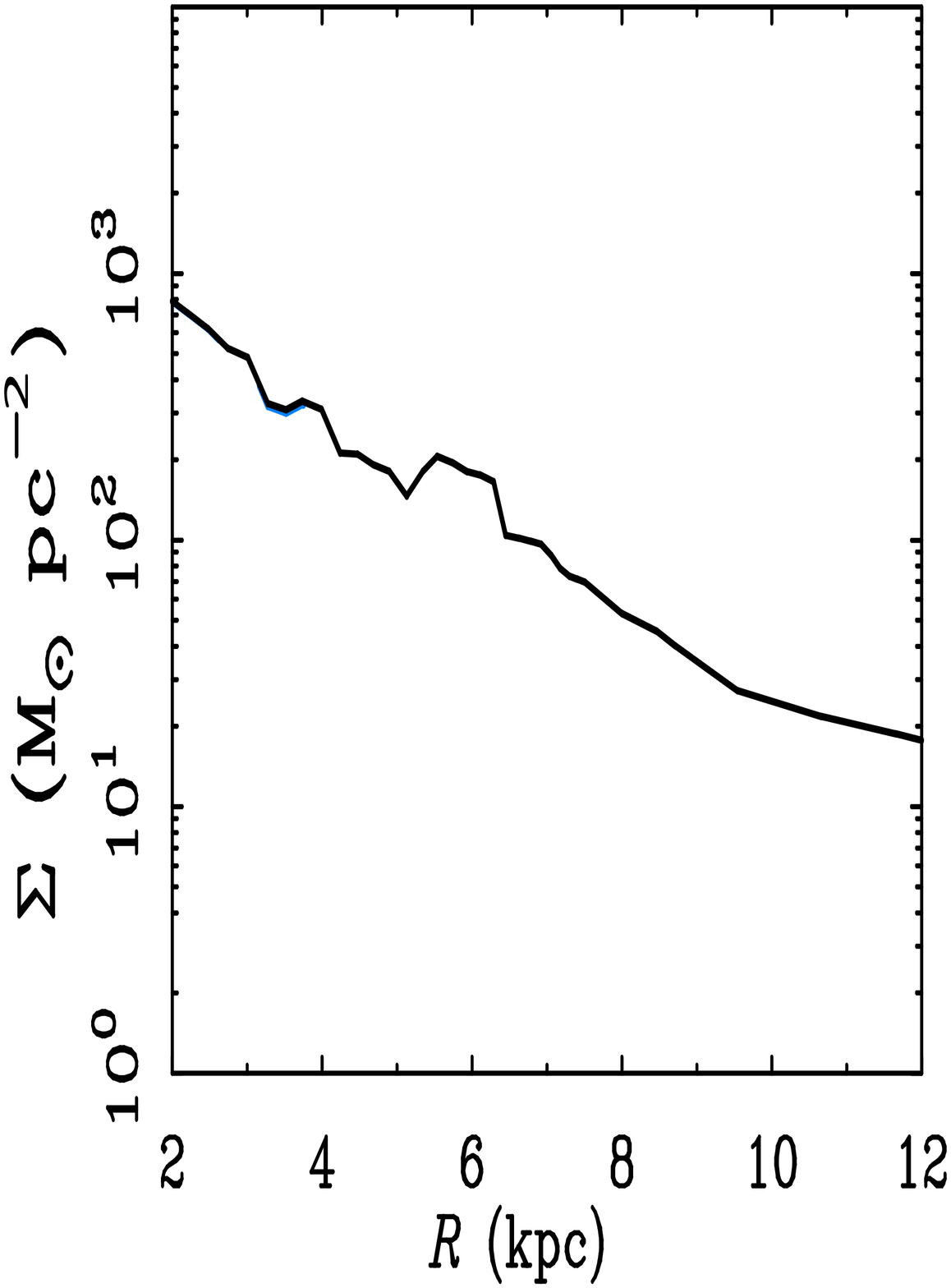}
\end{center}

Figure 14 The SMD of the Milky Way between $8$ and $12\  kpc$ (from figure 29 of Famaey and Mc Gaugh, 2012)

\begin{center}
\includegraphics[height=200pt, width=200pt]{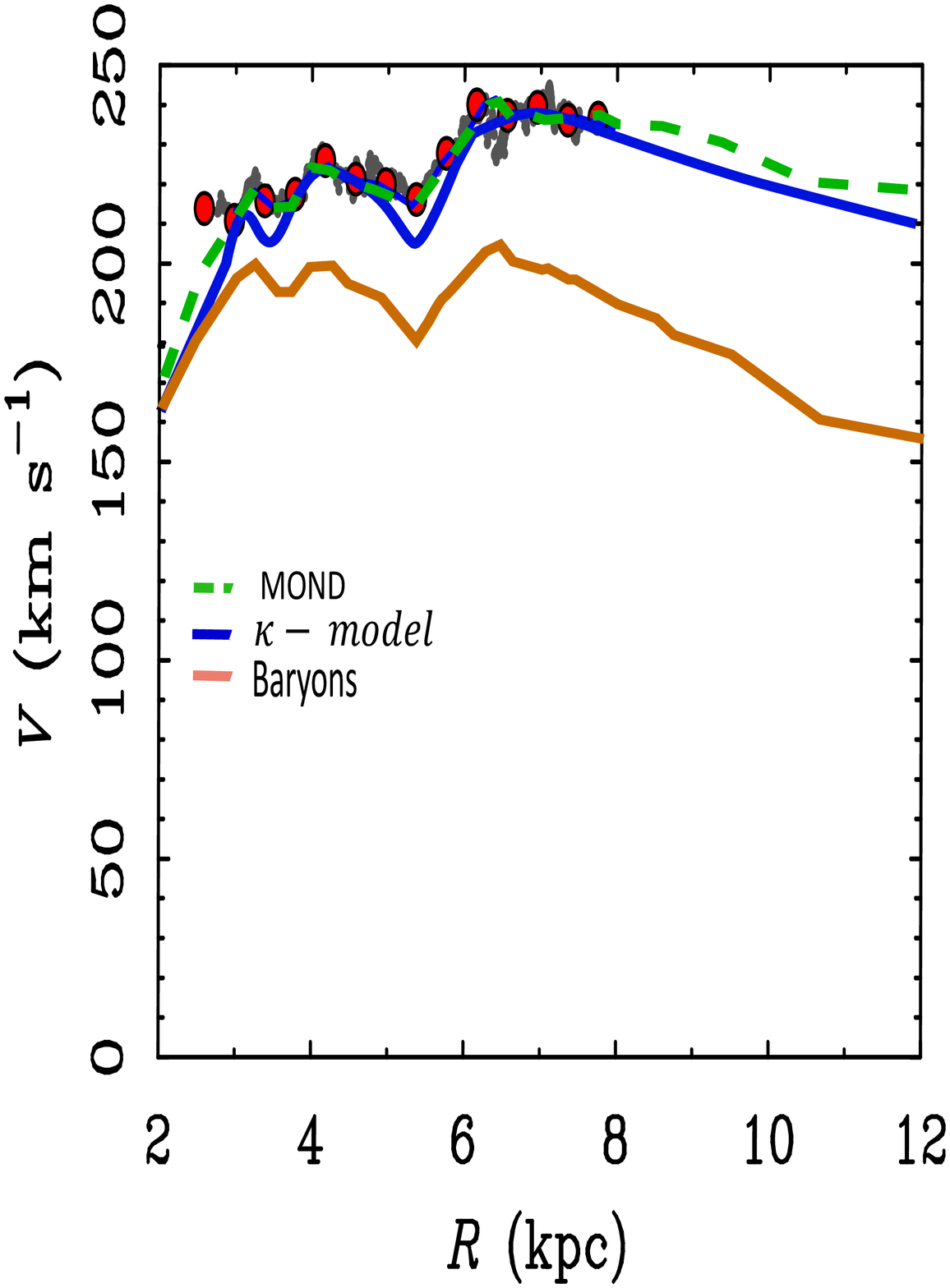}
\end{center}

Figure 15 The observed rotation curve of the Milky Way between $8$ and $12\  kpc$ (from figure 29 of Famaey and Mc Gaugh, 2012). The MOND curve is represented by the green dashed line, and the profile issued from the $\kappa$-model is displayed by the blue line.

\begin{itemize}
	\item \textbf{The SMD profile for the distances $0<r<100\ kpc$}
\end{itemize}

The  SMD profile for the distances   $0<r<100\ kpc$ is taken from Sofue (2020). It is reproduced in figure 16. With this SMD, the velocity of the Sun $v_{\bigodot}$, located  at 
$r_\odot=8\ kpc$ from the galactic center,  is   $\sim{}240\ kms^{-1}$ and the total
mass of the Miky Way without dark matter  (i.e. bulge + disk)  is $\sim{}\ 7\  {10}^{10}\ M_{\odot{}}$. The latter value is very close to that one  obtained elsewhere (Licquia and Newman, 2015).

\begin{center}
\includegraphics[height=130pt, width=250pt]{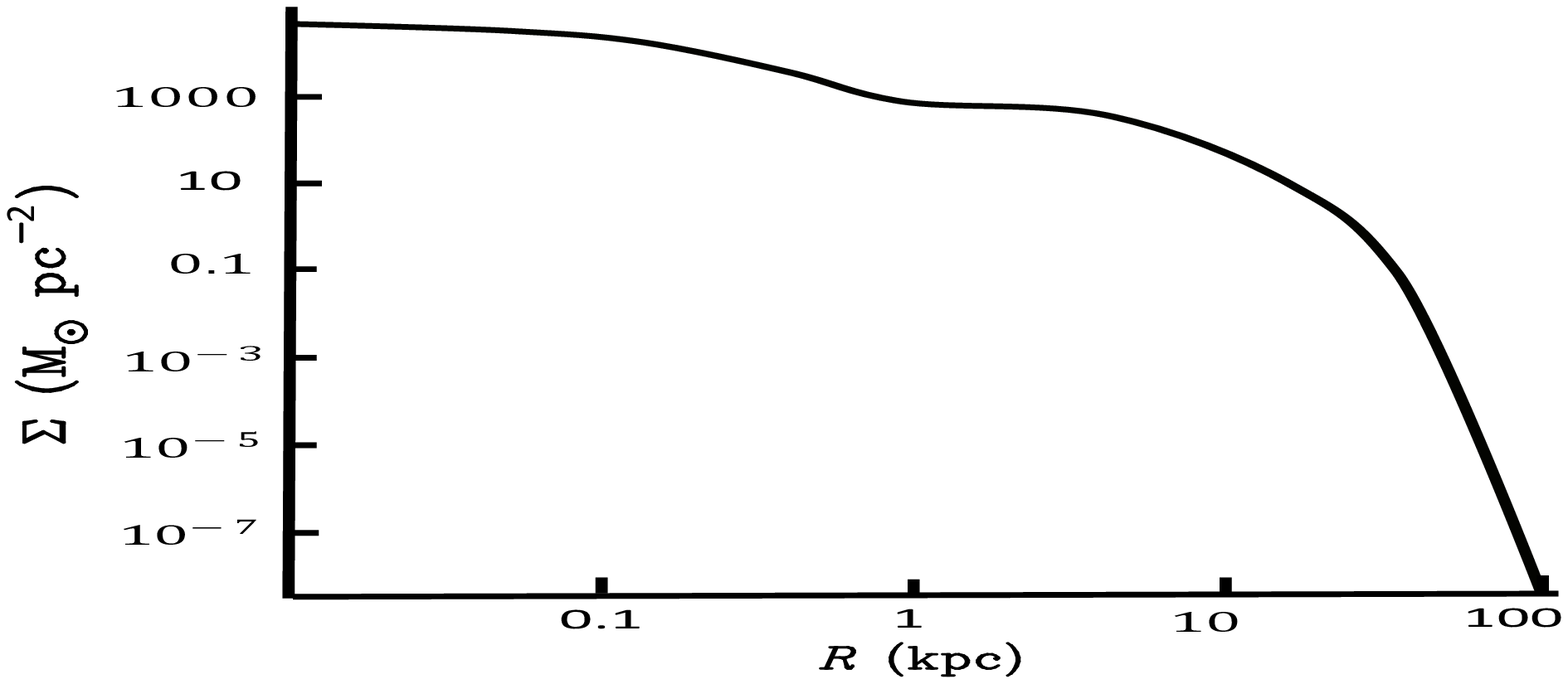}
\end{center}

Figure 16    The SMD of the Milky Way between $0$ and $100\  kpc$ obtained    from figure 3 of Sofue (2020) deprived of  the dark matter halo

\vspace{10pt}

The Newtonian profile in figures 17 and 18 is produced from the SMD (figure  16) following the same methodology as that described in Sofue (2020).

In  figure 17, the Newtonian and $\kappa{}-$ model curves are superimposed on
various observational curves obtained using various disk tracer
samples for $0<r<25\ kpc$ (figure 2 of Bhattacharjee, Chaudhury and Kundu, 2014). The $\kappa$-model curve is produced  by still using the same    global magnification factor (namely $0.46$) than in figure 15.
The observed rotation velocity shows obvious deviation from the predicted
Newtonian $r^{-1/2}$ law, but the $\kappa{}-$model curve fits the data, i.e., all  error bars overlap the $\kappa{}-$model curve. The error bars are, however, very large, which is linked to the difficulty of accurately estimating the distances and measuring the velocities from inside the Milky Way.

\begin{center}
\includegraphics[height=200pt, width=200pt]{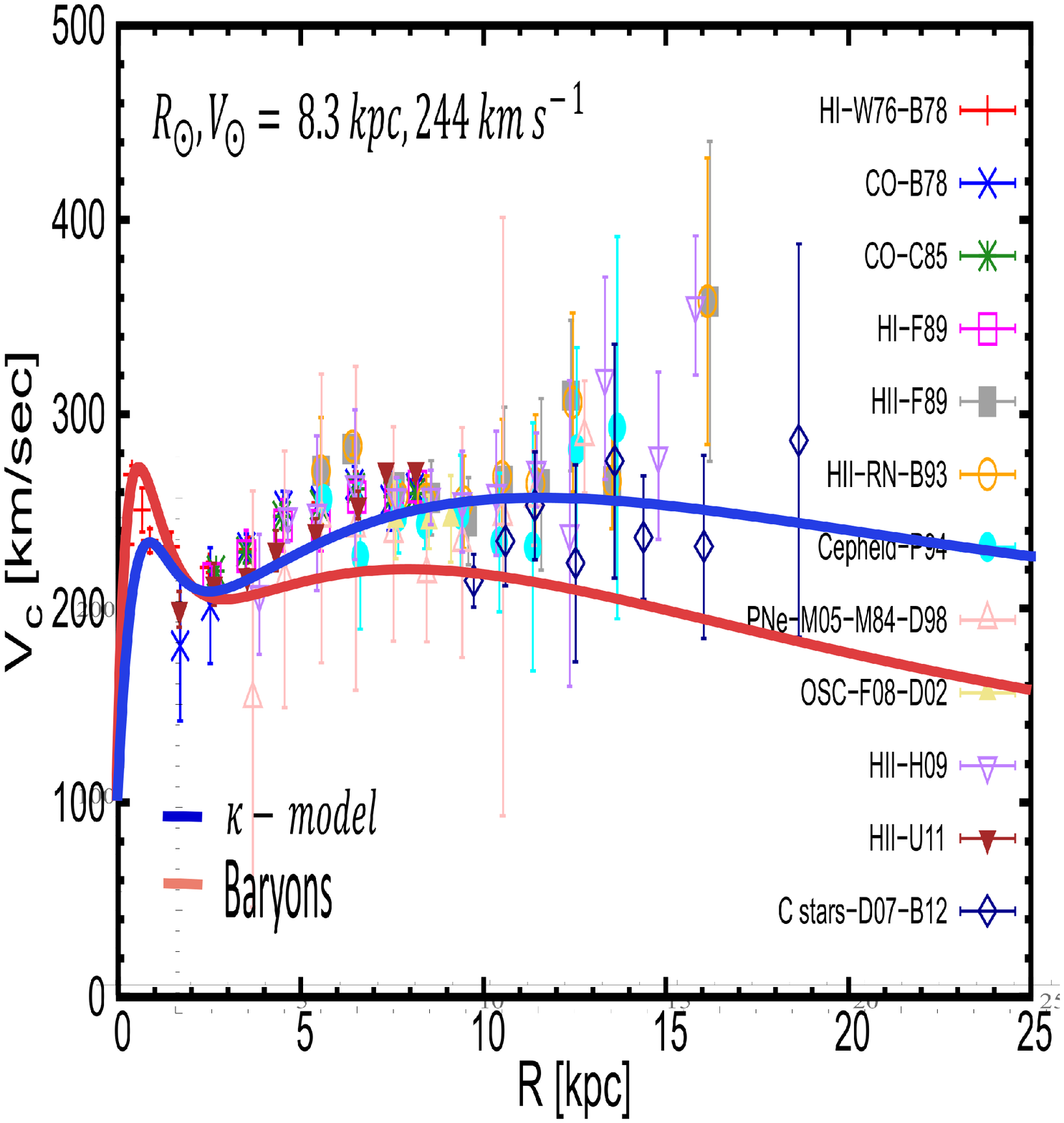}
\end{center}

Figure 17 A collection of data for the rotation curve of the Milky Way between $0$ and $25\ kpc$ obtained using various disk tracer samples (from figure 2 of Bhattacharjee, Chaudhury and Kundu, 2014). The $\kappa$-model curve is represented by the blue line.

\vspace{10pt}
  The domain
$0<r<100\ kpc$ is supplied in figure 18. This time, the Newton and
$\kappa{}-$model curves are superimposed on four dark matter profiles calculated
by Lin and Li (2019). Each of these dark matter profiles have two free
parameters. The observational data are taken from Huang et al. (2016). We can see
that the dark matter profiles and the $\kappa{}-$model profile  
fit equally the data, except around $20-40\ kpc$, where the observational results are
above all the curves. We can conclude that the rotation curve can be
reproduced in the framework of the $\kappa{}-$model, that is, an overall value of
$\sim{}210-250\ kms^{-1}$ for $4-20\ kpc$ (with a higher margin of error), beyond which the value declines steadily to
$175\ kms^{-1}$ at $100\ kpc.$ Obviously, we can claim that the same conclusion
is drawn from the dark matter models, but the key difference is that in the latter case, a conspiracy
is needed between the dark matter halo and the baryonic distribution of matter
to reproduce the quasi-flat part of the rotation curve.

\begin{center}
\includegraphics[height=200pt, width=250pt]{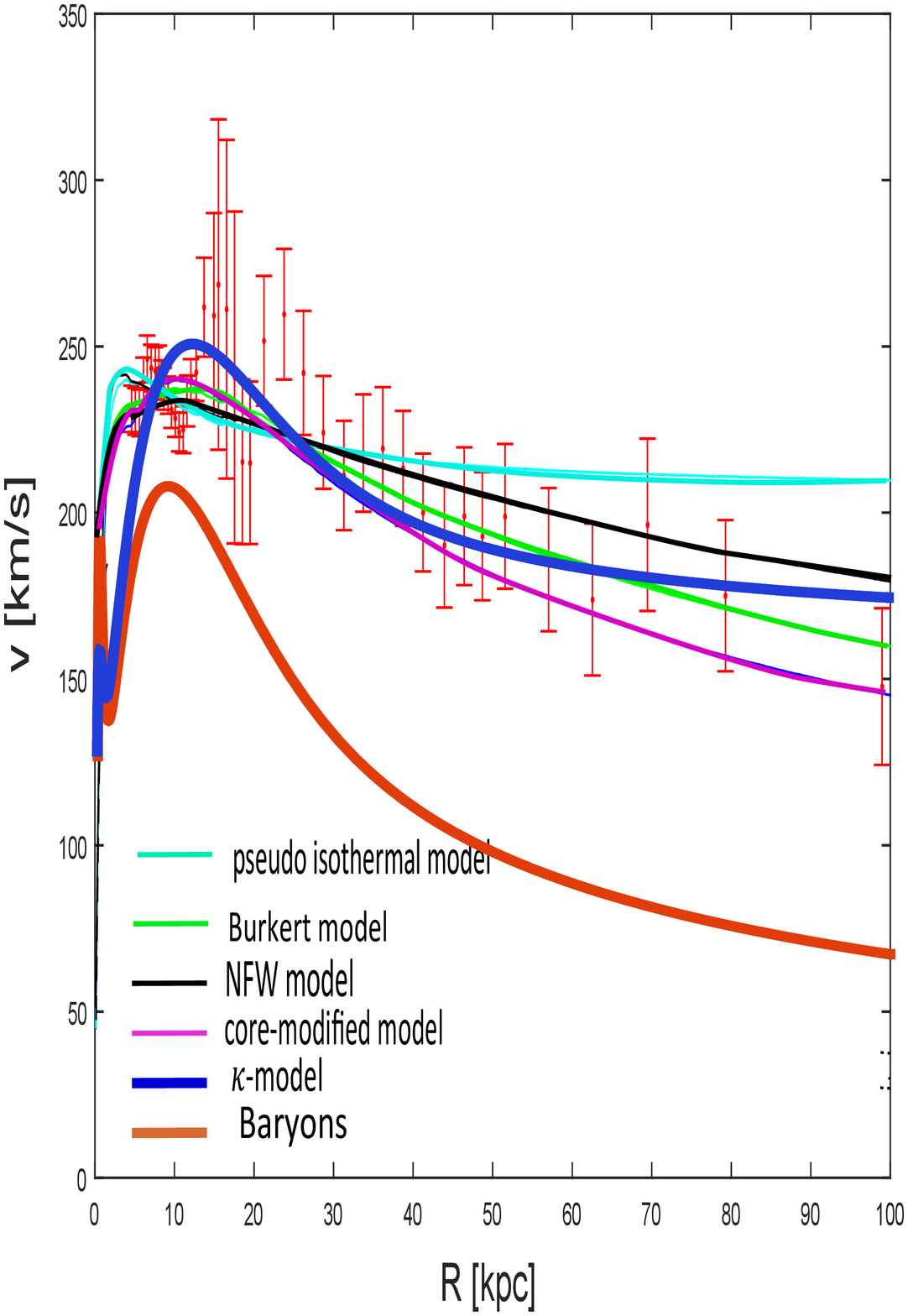}
\end{center}

Figure 18     The observational rotation curve of the Milky Way between $0$ and $100\ kpc$ compared to various dark matter profiles (from figure 2 of Lin and Li, 2019). The $\kappa$-model curve is represented by a thick blue line.

\vspace{10pt}
None of the theoretical profiles seen in figure 17 and figure 18 take into account the
spiral arms and rings, which also affect the rotation velocity. The presence of two prominent bumps on the rotation
curve has been reported (Huang et al., 2016). Let us note that these two bumps,
located at $12-13$ and $19-20\ kpc$, are not explained by the dark
matter halo (figure 12 of Huang et al, 2016). Huang et al. suggest that
two quite massive caustic rings of dark matter on the order of ${10}^{10}$
$M_\odot$ are needed to explain these bumps (Duffy and
Sikivie, 2008; Huang et al, 2016). Despite this interesting proposal, we cannot exclude the fact that these bumps could be simply
an artifact of the measurements. We may then wonder whether the addition of such
structures composed of dark matter is somewhat artificial. As 
acknowledged by Huang et al (2016), {\it{at present, the observational evidence linking
the dips seen in the rotation curve to hypothetical caustic rings of dark matter is still
marginal}}. Unfortunately, this type of methodology is typical of the dark
matter paradigm, which eventually suits all situations but which is
ultimately not really predictive. Nor should it be forgotten
that the rotation curve of the Milky Way still remains more poorly described than the
curves of other galaxies, where no such intriguing bumps are apparent. Thus,
Sofue (2020), in a review on the current status of
the study of rotation curve of the Milky Way, presents an SMD
distribution that is fitted by taking into account just the bulge, the disk composed of stars and gas,
and the Navarro-Frenk-White (NFW) dark halo with no mention of a hypothetical series of dark matter rings. Ultimately, the
$\kappa{}-$model is an even more economical solution, as it takes in
account only the baryonic matter; i.e., the Milky Way may be conveniently
parameterized by a bulge and a disk to produce an acceptable fit
of the observational curve as it is known today.

In conclusion, taking into account the various uncertainties regarding the measured velocities, the surface density profiles, the inclination of the galaxies and the estimate of the distances, the analysis of the individual examples performed clearly shows that with the sole consideration of the observed distribution of baryonic matter, both MOND and the $\kappa$-model can satisfactorily predict the observed rotation curves of galaxies. Such a prediction is not possible using the dark matter paradigm, unless we take a fine-tuned distribution of dark matter especially suited to each type of galaxy.

In addition, there is observational evidence that the peculiarities seen in the Newtonian curve at a location are nicely transcribed
to the observed curve at the same location, as attested in figures 10 and 15. This is a rather astonishing fact that has
already been pointed out by other authors (Famaey and McGaugh, 2012; McGaugh, 2016) and that is named Sancisi's law (Sancisi, 2004)\footnote{The mass discrepancy-acceleration relation (MDAR) formula (McGaugh, 2016) was a first attempt to quantify the phenomenon. Unfortunately, even though very noteworthy, this relation is purely empirical and is not based on a specific theoretical background.}. Here again, this phenomenon seems
to be very difficult to mimic using the dark matter paradigm, given that the gravitational force is a long-range force. Mass removal from a particular location does not create a trough in the observed velocity curve at the same location. An unlikely conspiracy between dark matter and baryonic mass is then needed for the trough in the Newtonian curve to be replicated on the observed velocity curve at the same place. This fact
is clearly favors models such as MOND and the $\kappa{}$-model, the
latter assuming a "simple" magnification of the velocity curve produced on the spot
from the baryonic model. Thus, following a pair of equations (6 and 7), this
magnification increases as the mean density (or, in an equivalent manner,
the factor $\kappa{}$) in the galaxy weakens. This circumstance is very well illustrated when comparing a high luminosity galaxy
(e.g., NGC 6946), where the magnification is relatively weak, and a low luminosity galaxy (e.g., NGC 1560), where the magnification is strong.

Ultimately, let us note that in spiral galaxies, the surface brightness profile typically varies in an exponential manner (Binney and Merrifield, 1998). Following the $\kappa$-model, this feature naturally implies a very extended radial range with near-constant rotation velocity. A remarkable example is Malin 1, which is a spiral galaxy exhibiting an extremely large low surface brightness disk of gas that is five times wider than the Milky Way (figure 19). Thus, Lelli, Fraternali, and Sancisi (2010) provide a flat rotation curve up to $100\  kpc$ for Malin 1 using HI data. However, Malin 1 is a relatively isolated galaxy, and it is very likely that the flat profile does not continue forever for any galaxy, especially when in the outermost regions of the disk, interaction with the environment, i.e., the influence of a neighboring galaxy, begins to be detected.

\begin{center}
\includegraphics[height=200pt, width=250pt]{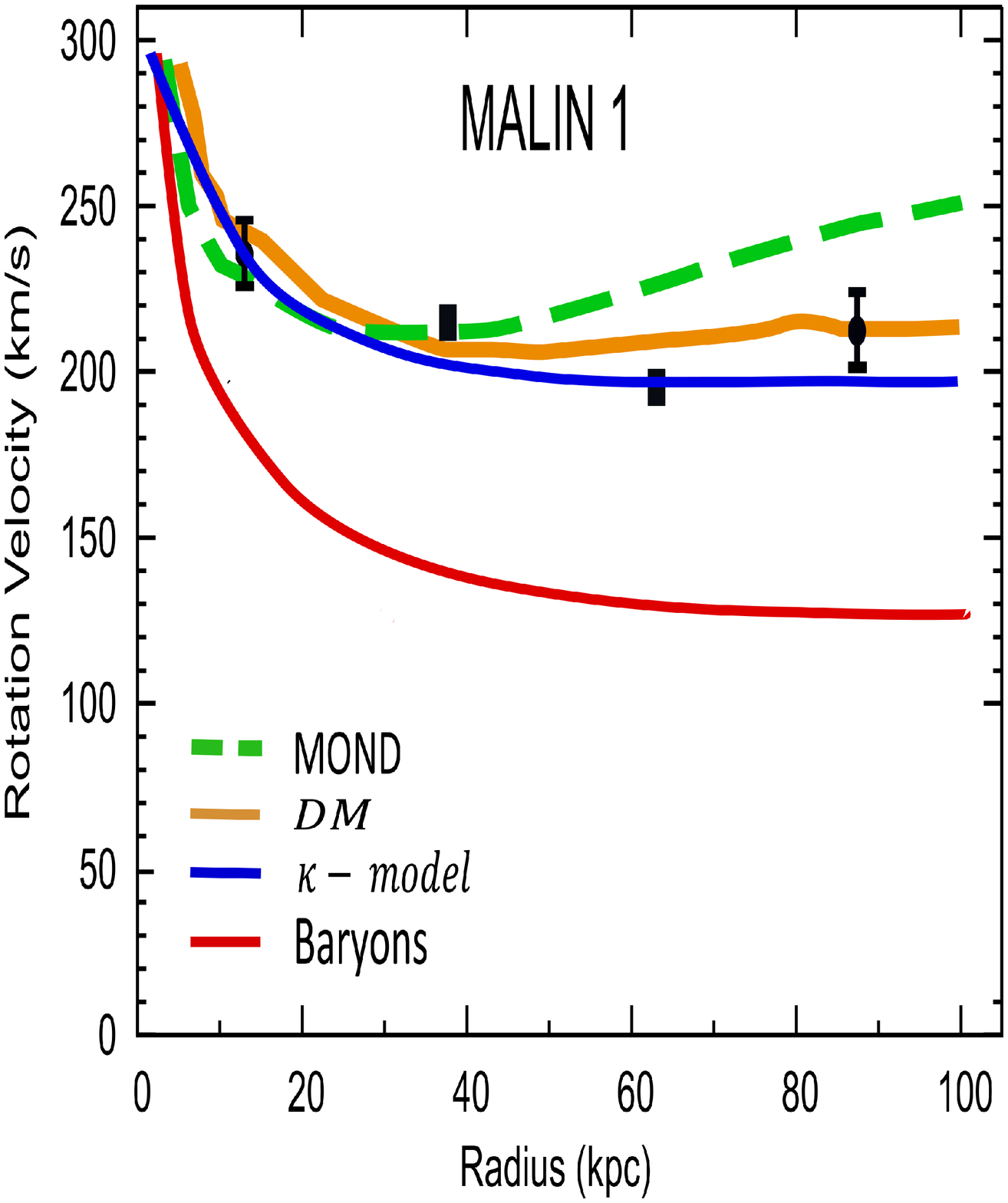}
\end{center}

Figure 19     The observed rotation curve of Malin 1 between $0$ and $100\ kpc$ compared to the dark matter, MOND and $\kappa$-model profiles (from figures 7 and 9 of Lelli, Fraternali and Sancisi, 2010). The $\kappa$-model curve is represented by a blue line.

\section{Conclusion}

The $\kappa$-model is a nice application of the notion of the - here apparent -
asymmetric distance well known in mathematics that essentially leads
to privileging local physics rather than global physics. Undoubtedly, the $\kappa$-model is no longer the whole of the story and might even ultimately be a pure
mathematical construction or a simple exercise of the mind. Nevertheless, regardless of these considerations, it should not be discarded too hastily, mainly due to the strong support of its conclusions, i.e., the
unification of some properties of galaxies under a sole umbrella, mainly an
unexpected, but however possible, correlation between the appearance of a
quasi-steady spiral substructure and the flat rotation curve in galaxies. Another
greatly interesting aspect of the $\kappa$-model is that it is not artificially parametrized and is sufficient by
itself, with the presence of only baryonic matter, which is a major issue.
In this sense, it is more easily refutable than any other theory postulating the
existence of a large variety of exotic particles with unknown characteristics that can be
changed at will.

An extension of this work is required. For elliptical galaxies and
globular clusters, a three-dimensional model is needed. However, a
forthcoming analysis of the data obtained for both the train
wreck and bullet clusters should enhance the understanding of the
density-dependent character of the measured gravitational force, which has been
assumed. Perhaps the $\kappa$-model could also help to eliminate the necessity of dark
energy. After the formation of the galaxies, the mean value of the factor $\kappa{}$ increases\footnote{The harmonic mean of $\kappa$ for a set of particles (a spiral galaxy, an elliptical galaxy or a cluster of galaxies) can be defined by $\frac{\kappa_{E}}{<\kappa>}=\sqrt{\frac{\int {\kappa_{E}^3}d^3{\sigma}\   \rho(\kappa_{E}{\boldsymbol{\sigma}})\  (\frac{\kappa_{E}}{\kappa})^2}{{\int {\kappa_{E}^3} {d^3{\sigma}}\  \rho({\kappa_{E}\boldsymbol{\sigma}})}}}$ The two integrations under the square root must be performed over the volume containing the set of particles.}. This phenomenon leads to a slow decline in the mean
gravitational forces between the galaxies, which could eventually be interpreted as an apparent acceleration of the Universe.

\vspace{10pt}
{\raggedright{\textbf{Data availability statement}: The author confirms that the data supporting the findings of this study are available within the article and the reference list.}}

\vspace{10pt}
{\raggedright{\textbf{Acknowledgements}:  I would like to thank the reviewer for his fruitful comments, which helped to improve the manuscript.}}

\vspace{10pt}
{\raggedright{
{\textbf{Conflicts of Interest:} The author declares no conflicts of interest.}}}

\section{References:}

{\raggedright{

Bhattacharjee, P., Chaudhury, S., \& Kundu, S., 2014, ApJ, 785

Benoit-Lévy, A., \& Chardin, G. 2012, A\&A, 537, A78

Binney, J.,  \& Merrifield, M., 1998, Galactic Astronomy, Princeton University Press

Brandao, C.S.S., \&  de Araujo, J.C.N., 2012,  ApJ,  750, 29

Capozziello, S., \& De Laurentis, M., 2012, Ann. D. Physik,  524, 545

Carignan, C., Charbonneau, P., Boulanger, F., \& Viallefond, F., 1990, A\&A, 234, 43

Conselice, C. J., 2018, Research Notes of the American Astronomical Society,
2, 43

Dobbs,  C.L., Theis, C., Pringle,  J.E., \&  Bate, M.R., 2010, MNRAS, 403, 625

Duffy, L.D., \&  Sikivie, P., 2008, Phys. Rev. D, 78, 063508

Elridge, J.J., \& Xiao,  L.,  MNRAS, 2019, 485, L58

Famaey, B., \& McGaugh, S.S., 2012, Living Rev. Relativity, 15, 10

Farnes,  J.S., 2018, A\&A, 620, A92

Frieman, J.A., Turner, M.S., \&  Dragan, H., 2008, Annual Review of Astronomy and Astrophysics, 46~(1), 385

Guedes, J.,  Callegari, S., Madau, P., \& Mayer, L., 2011, ApJ, 742, 76

Hobson, M., \& Lasenby, A., 2021, Phys. Rev. D 104, 064014

Huang, Y., Liu, X.W., Yuan, H.B., Xiang, M.S., Zhang, H.W., Chen, B.Q., Ren, J.J., Wang, C., Zhan,g Y., Hou, Y.H., Wang, Y.F.,  \&  Cao, Z.H., 2016, 463, 2623

Lelli, F., Fraternali, F. and Sancisi, R. 2010, Astronomy \& Astrophysics, 516, A11

McGaugh, S.S.,  Lelli, F., \&  Schombert J.M., 2016, Phys Rev Lett,  117, 201101

Licquia, T.C., \& Newman J.A., 2015, ApJ, 806, 96 

Lin, H.N., \& Li, X.,  2019, MNRAS, 487, 5679

Ludwig, G.O., 2021, Eur. Phys. J. C., 81, 186

 McGaugh, S.S.,  Lelli, F., \& Schombert, J.M., 2016,
Phys. Rev. Lett. 117, 201101

McGaugh, S.S., 2016, ApJ, 816, 42

Mancera Piña, P. E., Fraternali, F., Oosterloo, T., Adams, E. A. K., Oman, K. A., \& Leisman, L., 2021,  MNRAS. https://doi.org/10.1093/mnras/stab3491

Manfredi, G., Rouet, J.-L., Miller, B., \& Chardin, G. 2018, Phys. Rev. D, 98,
023514

Mannheim, P.D., \& Kazanas, D., 1989, ApJ,  342, 635

 Martinez-Medina, L.A.,  Bray, H.L., \& Matosa, T., 2015,  JCAP, 1512

Milgrom, M., 1983, ApJ,  270, 365

Milgrom, M., 2019, arXiv: 1910.04368v3

Moffat, J.W., 2006, J. Cosmol. Astropart. Phys., 2006, 4

 Moffat, J.W., 2008, Reinventing Gravity, Eds.
HarperCollins

Moni Bidin, C.,  Carraro, G.,  Méndez, R.A., \& R. Smith, R., 2012, ApJ, 751, 30

 O'Brien, J.G.,  \&  Moss, R.J.,  2015 J. Phys.: Conf. Ser. 615,  012002
 
 Pascoli,  G., \& Pernas,  L., 2020,  hal.archives-ouvertes.fr/hal-02530737
 
 Sancisi, R., 2004, The visible matter – dark matter coupling, in Ryder, S., Pisano, D., Walker, M.
and Freeman, K., eds., Dark Matter in Galaxies, IAU Symposium 220, p. 233, Astronomical Society of the Pacific, San Francisco
 
 S\'{a}nchez-Salcedo, F.J., \& Hidalgo-G\'{a}mez, A.M., 1999, A\&A,   345, 36
 
 Shu, F., 2016, Annual Review of Astronomy and Astrophysics, 54, 667

 Socas-Navarro, H., 2019, A\&A, 626, A5
 
 Sofue, Y.,  Honma M., \&  Omodaka  T., 2009, PASJ, 61, 227
 
 Sofue, Y., 2020, Galaxies, 8, 37
 
 Weinberg, S., 2008, Cosmology, Oxford University Press

\appendix
\section*{Appendix A}}
\renewcommand{\thesubsection}{\Alph{subsection}}

Here, we explain the origin of form (1) for the dynamics equation. We introduce the formal action

\vspace{-15pt}

\begin{equation}
S=\int dt\left[\frac{1}{2}m\left(\frac{{d{\bf{M}}}}{dt}\right)^2-V({M})\right]
\end{equation}

{\raggedright where $m$ is the mass of a test particle and $V(M)$ the potential experienced by this
particle located at a given point $M$. An arbitrary variation $\delta{}{\bf{M}}$ from $M$
gives}

\begin{equation}
\delta{}S=\delta{}\int dt\left[\frac{1}{2}m\left(\frac{{d{\bf{M}}}}{dt}\right)^2-V\left(M\right)\right]=\int dt\
\delta{}\left[\frac{1}{2}m{\left(\frac{{{d\bf{M}}}}{dt}\right)}^2-V\left(M\right)\right]
\end{equation}

\begin{equation}
=\int dt\
\left[m\frac{d{\bf{M}}}{dt}\delta{}\left(\frac{{d{\bf{M}}}}{dt}\right)-\delta{}V\left(M\right)\right]
\end{equation}

To continue the calculations, we must now exchange $d$ and $\delta{}$.

{\raggedright{
\textbf{The exchange of $d$ and $\delta{}$}}}

Let three observers $A$, $B$ and $C$ be located in a plane (figure
12). We can define this plane by imagining a common direction perpendicular to
${\bf{M}}_0{\bf{M}}_1$ and ${\bf{M}}_0{\bf{M}}_2$. This operation is possible because any orientation is well
defined in the $\kappa$-model. With the help of this figure, we write\footnote{Notation: The arrow $\longrightarrow{}$ indicates that a given
observer measures the corresponding bipoint. The symbol
$\triangleq{}$ indicates a definition, and the symbol $\equiv{}$ signifies that
the two compared vectors have the same orientation and the same norm but that each
of them is seen by a distinct observer.
}

\[
d{\bf{M}}\triangleq{}{\bf{M}}_{0}{\bf{M}}_1[d{\boldsymbol{\sigma{}}}]\vert{}_A\equiv{}d{\bf{M}}_{\parallel{}}={\bf{M}}_{2}{\bf{M}}_{1\parallel{}}[d{\boldsymbol{\sigma{}}}]\vert{}_{C}\longrightarrow{}\kappa{}d{\boldsymbol{\sigma{}}}  \  \ (a)
\]

The first expression signifies that the observer $A$ measures ${\bf{M}}_0{\bf{M}}_1$ and
obtains $\kappa{}d{\boldsymbol{\sigma{}}}$ and that observer $C$ measures
${\bf{M}}_2{\bf{M}}_{1\parallel{}}$ and obtains the same value. Other very similar relations
follow

\[
\delta{}{\bf{M}}\triangleq{}{\bf{M}}_0{\bf{M}}_2\left[\delta{}{\boldsymbol{\sigma{}}}\right]\vert{}_A\equiv{}{\delta{}{\bf{M}}}_{\parallel{}}\triangleq{}{{\bf{M}}_1{\bf{M}}_{2\parallel{}}\left[\delta{}{\boldsymbol{\sigma{}}}\right]\vert{}}_B\longrightarrow{}\kappa{}\delta{}{\boldsymbol{\sigma{}}}  \  \  (b)
\]

\vspace{-10pt}

\[
{{\bf{M}}_2{\bf{M}}_4^{"}\  \left[d{\boldsymbol{\sigma{}}}'\right]\vert{}}_C\longrightarrow{}(\kappa{}+\delta{}\kappa{})d{\boldsymbol{\sigma{}}}'\equiv{}{{\bf{M}}_2{\bf{M}}_4\left[d{\boldsymbol{\sigma{}}}+\delta{}d{\boldsymbol{\sigma{}}}\right]\vert{}}_A\longrightarrow{}\kappa{}\left(d{\boldsymbol{\sigma{}}}+\delta{}d{\boldsymbol{\sigma{}}}\right)
\  \ (a')\]

\vspace{-10pt}

\[
{\bf{M}}_1{\bf{M}}_4^{'}\  \left[\delta{}{{\boldsymbol{\sigma{}}}'} \right]\vert{}_B\longrightarrow{}(\kappa{}+d\kappa{})\delta{\boldsymbol{\sigma{}}}'\equiv{}{{\bf{M}}_1{\bf{M}}_4\left[\delta{}{\boldsymbol{\sigma{}}}+d\delta{}{\boldsymbol{\sigma{}}}\right]\vert{}}_A\longrightarrow{}\kappa{}\left(\delta{}{\boldsymbol{\sigma{}}}+d\delta{}{\boldsymbol{\sigma{}}}\right)
\  \  (b')\]

We must remark that ${d{\bf{M}}}_{\parallel{}}$ is $d{\bf{M}}$ parallelly displaced with respect to
itself respecting the conservation of the length during the transport
(likewise for ${\delta{}{\bf{M}}}_{\parallel{}}$  \textit{vs}  $\delta{}\bf{M}$). The
observer $C$ located at ${{{M}}}_2$ sees ${d{\bf{M}}}_{\parallel{}}$ exactly as 
observer $A$ located at ${M}_0$ sees $d{\bf{M}}$; thus, $d\bf{M}$ and ${d{\bf{M}}}_{\parallel{}}$
have the same orientation and the same norm but their origins are distinct.
However, these two vectors are perceived differently by observer $A$ located
at $M_0$. Figure 20 is the projection of the full set of vectors on the
background of this observer. Let us note that ${\bf{M}}_0{\bf{M}}_1$ is the real path and
${\bf{M}}_2{\bf{M}}_4^"$ is the corresponding varied path. Now, we set

\[
{d\delta{}{\bf{M}}\triangleq{}{\bf{M}}}_{2\parallel{}}{\bf{M}}_4^{'}\  \  \  \  \  \
{\delta{}d{\bf{M}}\triangleq{}{\bf{M}}}_{1\parallel{}}{\bf{M}}_4^{"}
\]

\begin{center}
\includegraphics[height=150pt, width=270pt]{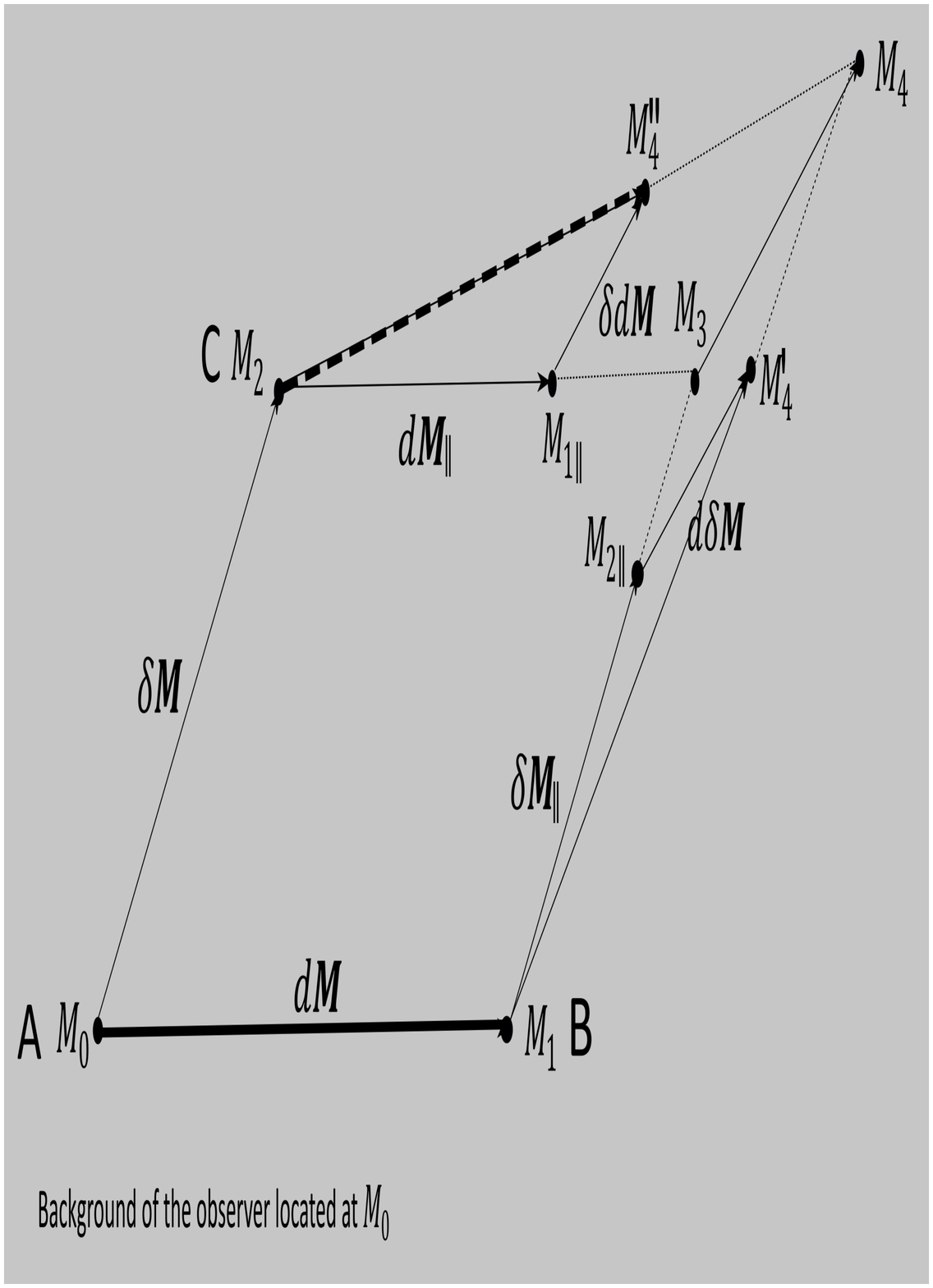}
\end{center}

Figure 20 Diagram of real and virtual paths: the thick full segment
represents the real path, the virtual path is indicated by a thick-dotted
line.

\vspace{10pt}

After subtraction $({a}')-(a)$ and $({b}')-(b)$, we have

\[
{\bf{M}}_1{\bf{M}}_4^{'}\  \left[\delta{}{{\boldsymbol{\sigma{}}}'} \right]\vert{}_B\longrightarrow{}(\kappa{}+d\kappa{})\delta{\boldsymbol{\sigma{}}}'\equiv{}{{\bf{M}}_1{\bf{M}}_4\left[\delta{}{\boldsymbol{\sigma{}}}+d\delta{}{\boldsymbol{\sigma{}}}\right]\vert{}}_A\longrightarrow{}\kappa{}\left(\delta{}{\boldsymbol{\sigma{}}}+d\delta{}{\boldsymbol{\sigma{}}}\right)
\  \  (b')\]

\[  {\bf{M}}_{1\parallel{}}{\bf{M}}_4^{"}   \left[{\delta{}d{\boldsymbol{\sigma{}}}'}\right]\vert{}_C\
\ \ \longrightarrow{}\kappa{}\delta{}d{\boldsymbol{\sigma{}}} \ \ \ \ \ \ \ \ \ \ \ \ \
{\bf{M}}_{2\parallel{}}{\bf{M}}_4^{'}  \left[d\delta{}{{\boldsymbol{\sigma{}}}'}\right]\vert{}_B\longrightarrow{}\kappa{}d\delta{
}{\boldsymbol{\sigma{}}}  \]

We naturally have $\delta{}d{\boldsymbol{\sigma{}}}=d\delta{}{\boldsymbol{\sigma{}}}$, thus 
(omitting the indexes)

\[
d\delta{}{\bf{M}}=\delta{}d{\bf{M}}
\]

Let us specify, however, that for observer $A$ located at $M_0$,
the vectors $d\delta{}{\bf{M}}$ and $\delta{}d{\bf{M}}$ (projected on the proper background) are
 parallel, but their (apparent) lengths are different, resp.
$\frac{\kappa{}}{\kappa{}+d\kappa{}}d\delta{}\boldsymbol{\sigma{}}$ and
$\frac{\kappa{}}{\kappa{}+\delta{}\kappa{}}\delta{}d\boldsymbol{\sigma{}}$; for
observer $A$, ${\bf{M}}_3{\bf{M}}_4[d\delta{}\boldsymbol{\sigma{}}]\longrightarrow{}\kappa{}d\delta{}\boldsymbol{\sigma{}}$, even
though these three vectors differ only by a small third order term. Let us
remark that
$\left(\kappa{}+\delta{}\kappa{}\right)d{\boldsymbol{\sigma{}}}'-\kappa{}d\boldsymbol{\sigma{}}=$
$\kappa{}\left(d{\boldsymbol{\sigma{}}}'-d{\boldsymbol{\sigma{}}}\right)+\delta{}\kappa{}d\boldsymbol{\sigma{}}'=\kappa{}\delta{}d{\boldsymbol{\sigma{}}}'+\delta{}\kappa{}d{\boldsymbol{\sigma{}}}'.$ This quantity is equal to $\kappa{}\delta{}d\boldsymbol{\sigma{}}$ (in
both orientation and norm), but
$\kappa{}\delta{}d{\boldsymbol{\sigma{}}}'+\delta{}\kappa{}d{\boldsymbol{\sigma{}}}'$ (origin
$M_{1\parallel{}}$) is evaluated by $C$, and $\kappa{}\delta{}d\boldsymbol{\sigma{}}\
$(origin $M_3$) is evaluated by $A$.}

\vspace{10pt}
After exchanging $d$ and $\delta{}$ in equation (10), we obtain

\[
\int dt\
\left[m\frac{d{\bf{M}}}{dt}\frac{d}{dt}\delta{}{\bf{M}}-{\boldsymbol{\nabla{}}}_M V\left(M\right)\delta{}{\bf{M}}\right]
\]

\begin{equation}
=\left.m\frac{{d{\bf{M}}}}{dt}\delta{}M\right\vert{}_{extremities}+\int dt\
\left[-\frac{d}{dt}\left(m\frac{{d{\bf{M}}}}{dt}\right)-{\nabla{}}_M V\left(M\right)\right]\delta{}M \  \  \
\end{equation}

For a stationary value of $S$, we have $\delta{}S=0$. We also take $\delta{}M=0$ at both extremities of the portion of the real trajectory. Eventually, we obtain

\begin{equation}
\frac{d}{dt}\left(m\frac{{d{\bf{M}}}}{dt}\right)+{{\boldsymbol{\nabla{}}}}_M V\left(M\right)={\bf{0}}
\  \  \  \ 
\end{equation}

This is the same expression as the usual dynamics equation in Newtonian mechanics. The
physics is left formally unchanged. This is very interesting. For
practical (computational) reasons, however, we rewrite this equation

\begin{equation}
\frac{d}{dt}\left(m\kappa{}\frac{d{\boldsymbol{\sigma{}}}}{dt}\right)+{\boldsymbol{\nabla{}}}_{(\kappa{}{\boldsymbol{\sigma{}})}}V\left(\kappa{}\boldsymbol{\sigma{}}\right)=0
\end{equation}

This is equation (1). The potential $V\left(M\right)$ is the gravitational potential. The dressed
potential, experienced by an observer located at a point $M$ and produced by
a point source (mass $\sf{M}$) located at an arbitrary origin (labeled by ${\boldsymbol{\sigma{}}}=\bf{0}$),
is\footnote{For two masses $m$ and ${m}'$ located at $M$ and ${M}'$, the
interaction potential is asymmetric and $V_M=-Gm{m}'\frac{1}{({\kappa{}}_M n
\Vert{}\sigma{}-{{\boldsymbol{\sigma{}}}}'\Vert{})}\ \not=\
V_{{M}'}=-Gmm'\frac{1}{({\kappa{}}_{{M}'}\
\Vert{}{\boldsymbol{\sigma{}}}-{{\boldsymbol{\sigma{}}}}'\Vert{})}$. The principle of
reciprocal action seems to be altered, but it must be kept in mind that the
two masses are not isolated and are not located in the same environment. Thus,
this fundamental principle is not violated, but it is not directly applicable
here. In contrast, the bare potential, $V_b=-Gm{m}'\frac{1}{
\Vert{}{\boldsymbol{{\boldsymbol{\sigma{}}}}}-\sigma{}'\Vert{}}$, is symmetric (even though it
is not measurable, given that $\Vert{}{\boldsymbol{\sigma{}}}-{\boldsymbol{\sigma{}}}'\Vert{}$  is
hidden.}

\begin{equation}
V\left({\kappa{}}_M{\boldsymbol{\sigma{}}}\right)=-G{\sf{M}}m\frac{1}{{(\kappa{}}_M{\boldsymbol{\sigma{}}})}
\end{equation}

We explicitly indexed the point where the potential is
measured by $\kappa{}$\footnote{Let us specify that we cannot directly measure a potential
(this quantity is defined up to a constant), but its gradient (the force on a
test particle of unit mass). However, this remark is simply a detail here.}. For a
shifted origin (at $O$), we likewise have

\begin{equation}
V\left({\kappa{}}_M\sigma{}\right)=-G{\sf{M}}m\frac{1}{({\kappa{}}_M\
\left\Vert{}{\boldsymbol{\sigma{}}}-{{\boldsymbol{\sigma{}}}}_O\right\Vert{})}
\end{equation}

Equations (14) and (15) need to be explained. The coefficient  ${\kappa{}}_M$, which
is linked to the observer located at $M$, determines the
intensity of the potential measured by this observer. It would
seem that it is the measurement process itself that imposes the distance that separates the observer from the attractive mass $\sf{M}$ on the observer. This circular
reasoning may sound irrational. In reality, both the measured distance and the
apparent gravitational potential felt by the observer depend on the mean
density $\bar{\rho{}}$ at $M$. It is the environment of the observer (and
obviously not the observer himself) that affects the measurements. There is
nothing strange about this. Thus, we can equivalently reason that by admitting that
 the attractive mass $\sf{M}$ is perceived by the observer as $\frac{\sf{M}}{{\kappa{}}_M}$, the
smaller ${\kappa{}}_M$ is, the higher the apparent
attractive mass and vice versa (however, the true mass is always $\sf{M}$).

\vspace{10pt}
\section*{Appendix B}

\textbf{Magnification formula}

We assimilate a galaxy into a steady and axisymmetric thin disk. The stars
travel in pure, uniform circular motion, and the coefficient $\kappa{}$ is independent of time. After removing the index $i$ for a test particle of
unit mass and taking   ${\kappa{}}_E=1$\footnote{The full equations with
the coefficient  ${\kappa{}}_E$ are

\[
{\frac{d}{dt}(\kappa{}}_E\frac{d{\boldsymbol{\sigma{}}}}{dt})={(\frac{{\kappa{}}_E}{\kappa{}})}^3\bf{F}_{New}
\]

{\raggedright and}

\[
{\bf{F}}_{New}=-Gm\sum_{j=1}^{N-1}\frac{{\kappa{}}_E({\boldsymbol{\sigma{}}}-{{\boldsymbol{\sigma{}}}_j})}{[{\kappa{}}_E\Vert{}{\boldsymbol{\sigma{}}}-{\boldsymbol{\sigma{}}]_j\Vert{}]}^3}
\]

}, can simplify equation (1) to

\begin{equation}
\frac{d}{dt}(\frac{d{\boldsymbol{\sigma{}}}}{dt})=\frac{{\bf{F}}_{New}}{{\kappa{}}^{3}}
\end{equation}

{\raggedright
where the Newtonian force ${\bf{F}}_{New}$ acting on the test particle is}

\begin{equation}
{\bf{F}}_{New}=-Gm\sum_{j=1}^{N-1}\frac{({\boldsymbol{\sigma{}}}-{\boldsymbol{\sigma{}}}_j)}{{\
\Vert{}{\boldsymbol{\sigma{}}}-{\boldsymbol{\sigma{}}}_j\Vert{}}^3}
\end{equation}

Because the trajectory of the test particle is circular and the force acting on it is purely radial, equation (16) immediately gives

\begin{equation}
\frac{{(\sigma{}\dot{\theta{}})}^2}{\sigma{}}=\frac{F_{New}}{{\kappa{}}^3}
\end{equation}

{\raggedright where $\theta{}$  designates the polar angle in the galactic plane from a
reference direction taken in this plane. This leads to}

\begin{equation}
\sigma{}\dot{\theta{}}=\frac{1}{{\kappa{}}^{\frac{3}{2}}}{{(F}_{New}\
\sigma{})}^{\frac{1}{2}}=\frac{1}{{\kappa{}}^{\frac{3}{2}}}{\ v}_{New}
\end{equation}

{\raggedright where ${v}_{New}$ is the Newtonian velocity. Eventually, the true
velocity\footnote{The true velocity refers to the local velocity that
is observable by spectroscopy.} $v$ is obtained by multiplying equation (19) by
$\kappa{}$. We obtain magnification formula 6 after reinserting
in it the coefficient ${\kappa{}}_E$.}

\end{document}